\documentclass[12pt]{article}

\usepackage{amsmath,amssymb,graphicx} 
\usepackage{pstricks}
\usepackage{cite}

\newcommand{\beq}{\begin{eqnarray}}
\newcommand{\eeq}{\end{eqnarray}}

\newcommand{\centeron}[2]{{\setbox0=\hbox{#1}\setbox1=\hbox{#2}\ifdim
\wd1>\wd0\kern.5\wd1\kern-.5\wd0\fi \copy0
\kern-.5\wd0\kern-.5\wd1\copy1\ifdim\wd0>\wd1
                                    \kern.5\wd0\kern-.5\wd1\fi}}
\newcommand{\ltap}{\>\centeron{\raise.35ex\hbox{$<$}}
                            {\lower.65ex\hbox{$\sim$}}\>}
\newcommand{\gtap}{\>\centeron{\raise.35ex\hbox{$>$}}
                            {\lower.65ex\hbox{$\sim$}}\>}

\newcommand\ZZ{\hbox{\zfont Z\kern-.4emZ}}
\font\zfont = cmss10 

\newcommand{\eref}[1]{eq.\ (\ref{e.#1})}
\newcommand{\erefn}[1]{ (\ref{e.#1})}

\newcommand{\aref}[1]{\ref{a.#1}}

\newcommand{\cref}[1]{Chapter \ref{c.#1}}

\def\nn{\nonumber \\}
\newcommand{\nl}{& \nonumber \\ &}

\def\beq{\begin{equation}}
\def\eeq{\end{equation}}
\newcommand{\ba}{\begin{array}}
\newcommand{\ea}{\end{array}}
\newcommand{\bea}{\begin{eqnarray}}
\newcommand{\eea}{\end{eqnarray} }
\newcommand{\bal}{\begin{align}}
\newcommand{\eal}{\end{align}}
\def\bi{\begin{itemize}}
\def\ei{\end{itemize}}
\def\ben{\begin{enumerate}}
\def\een{\end{enumerate}}
\def\beq{\begin{equation}}
\def\eeq{\end{equation}}
\def\bc{\begin{center}}
\def\ec{\end{center}}
\def\bt{\begin{table}}
\def\et{\end{table}}
\def\btb{\begin{tabular}}
\def\etb{\end{tabular}}
\newcommand{\bvec}{\left ( \ba{c}}
\newcommand{\evec}{\ea \right )}

\def\cl{{\mathcal L}}

\def\co{{\mathcal O}}

\def\gev{\, {\rm GeV}}
\def\tev{\, {\rm TeV}}

\def\mkk{\, M_{\rm KK}}

\def\mass2{mass${}^2$}

\def\ra{\rangle}
\def\la{\langle}

\def\pa{\partial}

\newcommand{\tr}{\mathrm T \mathrm r}

\def\simlt{\stackrel{<}{{}_\sim}}
\def\simgt{\stackrel{>}{{}_\sim}}

\newcommand{\ha}{{\hat a}}

\newcommand{\ti}{\tilde}

\def\ov{\overline}

\def\eps{\epsilon}

\begin{document}
\begin{titlepage}

\vskip1.5cm
\begin{center}
{\huge \bf Electroweak Breaking on a Soft Wall}
\vspace*{0.1cm}
\end{center}
\vskip0.2cm

\begin{center}
{\bf Adam Falkowski$^{a,b}$ and Manuel P\'erez-Victoria $^{c}$}

\end{center}
\vskip 8pt

\begin{center}
$^a$ { \it CERN Theory Division, CH-1211 Geneva 23, Switzerland} \\
\vspace*{0.3cm}
$^b$ {\it Institute of Theoretical Physics, Warsaw University,
            \\ Ho\.za 69, 00-681 Warsaw, Poland } \\
\vspace*{0.3cm}
$^c$ {\it CAFPE and Departamento de F\'{\i}sica Te\'orica y del Cosmos, \\ 
             Universidad de Granada, E-18071, Spain } 

\vspace*{0.3cm}

{\tt adam.falkowski@cern.ch, mpv@ugr.es}
\end{center}

\vglue 0.3truecm

\begin{abstract}
\vskip 3pt \noindent 

We extend the 5D gauge-higgs scenario to the soft-wall framework where the IR brane is replaced by a smoothly decaying warp factor.
The electroweak symmetry of the Standard Model is embedded in a larger symmetry group whose gauge bosons propagate in the bulk of a warped fifth dimension. 
This gauge symmetry is partly broken by UV boundary conditions and by a condensate of a bulk scalar field. 
The Higgs boson lives partly in the 5th component of the gauge field, and partly in the  the bulk scalar.   
The Higgs potential is not UV sensitive if the condensate vanishes fast enough in the UV region. 
The soft-wall realization opens new possibilities for the spectrum and the couplings of the Kaluza-Klein resonances.
We study two particular soft-wall backgrounds: one with resonances whose masses follow the linear Regge trajectory, and another with a continuum above a mass gap.  
We find that constraints on the Kaluza-Klein scale from electroweak precision tests are less severe than in analogous models with the IR brane. 
For the linear spectrum the typical constraint on the lightest resonance mass is 2 TeV, while the continuum is allowed to start below 1 TeV.

\end{abstract}

\end{titlepage}



\section{Introduction}
\label{sec:i} \setcounter{equation}{0} \setcounter{footnote}{0}

Warped extra dimensions occupy an important place in particle physics today. 
The original motivation of Randall and Sundrum was to address the hierarchy problem by lowering the cut-off scale in the Higgs sector \cite{RS}.
After several improvements, of which the most important  are placing the gauge and fermion fields in the bulk \cite{P} and extending the Standard Model (SM) gauge symmetry so as to include the custodial symmetry \cite{ADMS}, the Randall-Sundrum (RS) set-up remains a valid framework for studying electroweak symmetry breaking.    
Yet the most exciting aspect of RS is that, according to the AdS/CFT correspondence~\cite{adscft}, it provides a rough description of a purely 4D system where fundamental fields interact with a large N strongly-coupled, approximately conformal sector (CFT) \cite{APR}.
In particular, RS with a Higgs boson localized on the IR brane is a dual realization of the old idea of a composite Higgs boson arising as a bound state of some new strong interactions. 
Moreover, a 4D composite pseudo-Goldstone boson \cite{GK} can be modeled in RS by the so-called gauge-higgs scenario \cite{M}, where the Higgs boson is identified with the fifth component of a 5D gauge boson \cite{CNP}.  
Confidence in this 4D/5D correspondence is strengthened by the success of the related AdS/QCD approach in modeling the low-energy meson sector of QCD \cite{EKSS,DP}. 

The warped 5th dimension of RS is an interval terminated by the UV brane and the IR brane. 
From the holographic point of view, the fields living on the UV brane are interpreted as the fundamental sector probing the CFT, while the IR brane describes the low-energy dynamics of the CFT that spontaneously breaks the conformal symmetry and, often, other global symmetries.   The {\em hard-wall} IR brane is an idealization that translates to breaking the symmetries by a vev of an operator of infinite scaling dimension.   
It is interesting to investigate a more general set-up where the IR breaking is modeled by a smooth evolution of the background geometry and/or bulk fields vevs.
Such framework is referred to as  {\em the soft wall}.   

Soft walls have a relatively short and not so intense history.
The apparent reason is that the soft-wall backgrounds are inevitably more complicated than RS with a metric being a slice of AdS.  
The studies so far have been restricted to AdS/QCD, with the motivation of constructing a more faithful dual description of the QCD dynamics. 
The interest was spawned by the observation in ref. \cite{KKSS} that a carefully designed soft-wall background leads to a linear Kaluza-Klein (KK) spectrum: $m_n^2 \sim n$, rather than $m_n^2 \sim n^2$ as encountered in the hard-wall RS.
The linear spectrum of excited vector mesons in QCD is both expected by theoretical arguments and observed experimentally.   
Refs. \cite{CR,GKN} discussed more extensively how features of the soft-wall background map onto the properties of QCD. 
A dynamical system that leads to the background with linear KK trajectories has been proposed in ref. \cite{BG}. 

In this paper we investigate the soft wall in the context of electroweak symmetry breaking. 
The most obvious possibility would be to use a bulk scalar to break the electroweak symmetry. 
A working model can be constructed along the same lines as in AdS/QCD with $SU(2)_L \times SU(2)_R$ gauge group broken to $SU(2)_V$ by a bulk Higgs field in the bi-fundamental representation. 
Then the ``pions'' that result from this symmetry breaking pattern play the role of the electroweak would-be-Goldstones eaten by the W and Z bosons.
The difference with respect to AdS/QCD is that the $SU(2)_L \times U(1)_Y$ subgroup of the bulk gauge group should remain unbroken on the UV brane and that an additional $U(1)$ gauge factor has to be included in the bulk to correctly accommodate the hypercharge. 
Such a 5D set-up would be dual to a composite Higgs arising from a technicolor-like strongly coupled dynamics.
In this paper we jump immediately to a higher level where the Higgs is realized as a pseudo-Goldstone boson, that is to the 5D gauge-higgs scenario. 
This has the attractive feature that the Higgs is proteced by approximate global symmetries and therefore it can naturally be light.    

The generalization of the gauge-higgs models to the soft-wall case is not so straightforward, 
as there is no IR brane to break the bulk gauge symmetries. 
Instead, we have to introduce a charged bulk scalar field with a potential that forces it to obtain a vev.
The consequence is that the SM higgs boson lives not only in the fifth component of the broken gauge bosons, 
but also in the Goldstone bosons hosted by the bulk scalar.   
That implies that the radiatively generated Higgs mass can be UV sensitive because the scalar mass term is not protected by the symmetries at the UV brane.
We will show however that the UV sensitivity can be avoided if the bulk condensate is localized in IR and decays fast enough in the vicinity of the UV brane.

The hope is that the soft-wall version of RS provides a more adequate description of the dual composite Higgs models. 
It also offers more possibilities for  the KK spectrum and the couplings which is important in the context of LHC search strategies.   
One important thing we show in this paper is that the constraints on the KK scale from electroweak precision tests turn out to be somewhat milder.
Even small differences in allowed KK masses are extremely relevant from the phenomenological point of view, since they may greatly increase the prospects for a discovery at the LHC.
We construct an explicit soft-wall example where the typical constraint on the lightest KK mode mass is reduced down to 2 TeV, rather than 3 TeV typically found in hard-wall models.
Our another, more exotic example with a continuous KK spectrum separated from the SM particles by a mass gap is even less constrained and admits the continuum below 1 TeV.

In Section~2 we discuss general features of the soft wall scenario.  
We systematize various possibilities for the KK spectrum.
We also derive important properties of the solution to the equations of motion. 
This section is a bit loaded with math but the results are very relevant for phenomenological applications. 
Next, we move to the soft-wall version of the gauge-higgs scenario.   
In Section~3 we discuss a toy model based on the U(1) gauge symmetry in the bulk. 
This simple set-up allows us to understand the physics of the gauge-higgs and identify all degrees of freedom that arise in the soft wall set-up. 
KK scalars and pseudoscalar always appear on a soft wall (while they are optional in the hard wall version), and we devote some time to discussing their equations of motions.   
Section~4 is the heart of this paper. 
We consider a soft-wall version of the 5D gauge-higgs model based on SO(5) gauge symmetry, which is an example of a  fully realistic and calculable model of the electroweak sector.   
We discuss the spectrum of the gauge bosons and evaluate the gauge contribution to the radiative Higgs potential.  
We comment how the softness of the loop corrections in the usual hard-wall scenario can be maintained in the soft-wall version.  
Then we derive the low-energy action and general expressions for the electroweak precision parameters. 
In Section~5 we examine the electroweak sector in two particular soft-wall backgrounds. 
One has a discrete resonance spectrum, which however shows a different spacing between KK modes, as compared to the hard wall models. 
The other has a continuous spectrum above a mass gap, which could never be obtained with a hard wall.      
We analyze contraints from electroweak precision data in both scenarios and point out that they are less severe than in typical RS scenarios. We conclude in Section~6. Finally, an appendix contains the derivation of the effective action that we use to calculate the oblique parameters.

\section{Soft-Wall Background}
\label{sec:swb} \setcounter{equation}{0} \setcounter{footnote}{0}
We consider a 5D gauge theory propagating in a warped background with the line element   
\beq
\label{e.wb}
ds^2 = a^2(z) (dx_\mu^2 - dz^2). 
\eeq 
The metric in \eref{wb} may refer to the true background metric that solves the 5D Einstein equation, 
or it may be an effective background that incorporates a vev of some dilaton field multiplying the gauge action: $a_{eff}(z) = e^{\Phi(z)} a_{\rm true}(z)$ .
In this paper we are not concerned with the dynamics that could produce a particular warp factor.
In other words, we study 5D gauge theories in a fixed, non-dynamical background.  
This way we sweep under the carpet such important questions as radion stabilization, backreaction of condensates on the geometry,  etc.  
A proper discussion of these issues could easily obscure our main point which is low energy phenomenology.     
Therefore, we adopt a pragmatic approach and concentrate on the effect of general backgrounds on the observables in the electroweak sector.    See refs. \cite{CR,GKN,BG} for discussion of possible dynamical origins of the soft-wall background.

The conformal coordinate $z$ runs from $z_0$ (the UV brane) to infinity. 
We fix $a(z_0) = 1$. The strong-coupling cuf-off scale on the UV brane is then set by $\sim 4\pi/z_0$.   
Even though there is no IR brane, the KK spectrum develops a mass gap if the warp factor decays exponentially or faster in IR. 
If this is the case, the proper length of the extra dimension, $L = \int_{z_0}^{\infty} a(z)$, is also finite. %
Note that the finite proper length implies that in the coordinates $ds^2 = a^2(y) dx^2 - dy^2$ the 5th coordinate spans a finite interval $y \in [0,L]$. 
The difference with the usual RS scenario would be the vanishing of the warp factor on the "IR brane".     

The warp factor is not specified in most of the following discussion, and we only make a couple of  technical assumptions. 
One is that $a(z)$ is monotonically decreasing ($a' < 0$) for all $z$; the other is that it decays sufficiently fast at large $z$, so as to generate a mass gap.  
We also tacitly assume that the metric approaches AdS close to the UV brane, but our results apply to more general cases as well.  

A gauge field propagating in the background of \eref{wb} has a quadratic action of the form 
\beq
\label{e.5da}
S_5 = \int d^4 x \, dz  \, a(z) \left ( -{1 \over 4} F_{MN}^2 + {1 \over 2} M^2(z) A_M^2 \right ) . 
\eeq  
The mass term should be understood as resulting from a condensation of a charged scalar field and, in general, it can have a non-trivial dependence on the $z$ coordinate. 
This implies the condition $M^2(z) \geq 0$ for all $z$. 
The equation of motion that follows from the 5D action is   
\beq
\label{e.geom}
\left (a^{-1} \pa_z (a \pa_z )  - M^2 + p^2 \right ) f = 0.  
\eeq 
We are interested in the normalizable solution of this equation, 
in the sense $\int_{z_0}^\infty a f^2 < \infty$, and we  denote such a solution by $K_M(z,p)$ (when a normalizable solution does not exist, as in AdS, we define $K_M$ as the solution that exponentially decays for large Euclidean momenta).  
In the soft-wall set-up, normalizability plays the same role as IR boundary conditions in RS, selecting one of the two independent solutions of \eref{geom}. 
Then the UV boundary conditions leads to  a quantization condition for $p^2$, which fixes the KK spectrum.  

In the following of this section, we discuss general properties of the solutions to \eref{geom}.  
The formal results that we obtain here will later prove valuable to study physical observables in realistic models.  

To proceed, it is convenient to borrow some methods and terminology from quantum mechanics. 
The equation of motion can be recast into the form resembling the Schr\"odinger equation by defining the "wave function" $\Psi$ as $f = a^{-1/2} \Psi$.  
Note that the normalization condition translates to square-integrability: $\int \Psi^2 < \infty$.
The wave function satisfies  
\beq
\left ( - \pa_z^2 + V_M(z)  \right ) \Psi = p^2 \Psi \, , 
\qquad
V_M(z) = M^2 + {a''\over 2 a} - {(a')^2 \over 4 a^2}. 
\eeq 
From the shape of the potential one can quickly infer that the existence of the mass gap relies on the corresponding Schr\"odinger potential $V_M$ being confining. 
The necessary condition  reads $V_M \geq  {\rm const} > 0$ for  $z \to \infty$.  
Moreover, in order to have a minimum of the potential we also need $V_M$ to grow toward UV. 
This last condition is always fulfilled by metrics that are asymptotically AdS in UV, in which case $V_M \sim 1/z^2$ at small $z$. 

Furthermore, it is profitable to introduce the so-called  superpotential $W(z)$ \cite{SQM}, that is related to the Schr\"odinger potential by the Riccati equation 
\beq
\label{e.sde} 
W^2 - W' = V_M .
\eeq
One can prove that the asymptotic condition $V_M \geq {\rm const} > 0$ is equivalent to  $W \geq {\rm const} > 0$. 
This is obvious when $W^2$ dominates over $W'$. 
On the other hand, it is not possible to keep $W' > W^2$ asymptotically: sooner or later $W^2$ will catch up, bringing us back to the previous case. 

Embedding the SM electroweak sector in a 5D gauge theory will require that some of the bulk gauge symmetries remain unbroken, which translates to $M^2=0$ in the equation of motion for the corresponding generator.   
Therefore, we are interested in the backgrounds that have a mass gap in the limit $M^2 \to 0$, 
that is  $V_0$ must be confining (which implies that $V_M$ is confining too, as long as $M^2$ is positive). 
For $M^2 = 0$ the superpotential is $W_0  = - {a' \over 2 a} > 0$.
This shows that the KK spectrum in the unbroken phase has a mass gap if the warp factor decays in IR at least as fast as $e^{- z^\alpha}$ with $\alpha \geq 1$ \cite{GKN}.    
Depending on the power $\alpha$, three general situations can arise: 
\ben
\item {\bf Unparticles}, $W_0(z)|_{z \to \infty} \to 0$. 
The spectrum consists of a continuum of non-normalizable modes and there is no mass gap \cite{G}. 
The familiar example is that with the AdS metric $a(z) = z_0/z$ corresponding to   $W_0 = 1/2z$, which is nothing but the RS2 set-up \cite{RS2}. 
This direction is not explored in this paper.   
\item {\bf Hidden Valley}, $W_0(z)|_{z \to \infty} \to \rho > 0$.
The spectrum again has a  continuum of non-normalizable modes separated from the (optional) massless mode by a mass gap $\rho$ \cite{SZ}. 
An example presented recently in  ref. \cite{CMT} has $a(z) = e^{- 2\rho z}/z$ corresponding to $W_0(z) = 1/2z + \rho$.  
\item {\bf Resonances}, $W_0(z)|_{z \to \infty} \to \infty$. 
This is the most familiar scenario. 
The spectrum consists of a discrete tower of vector resonances separated by a mass gap from (optional) zero modes.
An important example based on the proposal of ref. \cite{KKSS} has the metric  $a(z) = e^{- \rho^2 z^2}/z$ corresponding to $W_0(z) = 1/2z + \rho^2 z $.  
\een 
One can also envisage hybrid scenarios where unbroken gauge boson are of the unparticle or the hidden-valley type, 
while broken gauge bosons have a discrete spectrum due to $M^2(z)$ asymptotically growing in IR.  

\vspace{.5cm}

We first study the normalizable solution to the equation of motion \erefn{geom} in the special case $p^2 = 0$: 
\beq 
\label{e.ee}
\left [ a^{-1}\pa_z (a \pa_z) - M^2(z) \right ] \eta   = 0 .
\eeq   
We will see in the next section that $\eta(z) \equiv K_M(z,0)$ is directly related to the gauge-higgs profile in the 5th dimension.  
Obviously, for $M^2 = 0$, the normalizable solution is just $\eta = 1$. 
In the following we concentrate on the case  $M^2 > 0$.   
To get more insight into the shape of $\eta$ we split  the superpotential as 
$W = W_0 + U_M$ where $W_0 = - a'/2a$, and $U_M$ is related to $M^2$ by the non-linear equation  
\beq
\label{e.mse}
U_M^2 - U_M' - {a'\over a} U_M = M^2. 
\eeq
We also fix $U_M(z_0) > 0$. 
Eq. \erefn{ee} can now be written as   
\beq
a^{-1} \left ( \pa_z - U_M \right )a \left ( \pa_z + U_M \right ) \eta(z) = 0.    
\eeq 
The normalizable solution is  $\eta(z) =  e^{- \int^z U_M}$. 
Normalizability follows from  the fact that $a \eta^2 = e^{- \int^z W}$ (recall that $\lim_{z\to \infty} W > 0$ is the mass gap condition). 

We will prove now that $\eta(z)$ is monotonically decreasing all the way from the UV brane down to IR. 
The derivative is  
$\pa_z \eta = -   U_M (z) e^{ - \int^z U_M(z') }$. 
Since $U_M(z_0) > 0$, the gauge-higgs profile decreases in the vicinity of the UV brane.
This trend could be reversed if $\eta$ had an extremum, which would imply that $U_M$ vanishes somewhere,   
$U_M(z_*) = 0$. 
Note however that $U_M$ must always decrease in the vicinity of $z_*$ since, 
from the equation \erefn{mse}, $- U_M'(z_*) = M^2(z_*) > 0$.  
Thus, $U_M$ could have at most one zero, if $U_M$ started positive at $z_0$ and became negative asymptotically. 
That is however incompatible with the asymptotic $\lim_{z \to \infty} W (z) > 0$.
Indeed, suppose $W > 0$ which implies $W_0 > |U_M|$.
From \eref{mse}, 
\beq
0 < M^2  = |U_M|^2 - 2 W_0 |U_M| - U_M' <  - W_0 |U_M|  - U_M'.
\eeq    
This requires that $U_M'$ be negative ($U_M$ is decreasing) and that $|U_M'| > W_0 |U_M|$.
Thus $\pa_z \log |U_M|  > W_0$: the logarithm of $U_M$ has to be steep enough such that it is larger than $W_0$.
But then it is impossible to keep $|U_M|  < W_0$ all the way down to IR.
We conclude that, as long as there is a mass gap, $U_M$ can never become negative and that {\em $\eta (z)$ is always decreasing.} 
This is an important result that will later turn out to be equivalent to positivity of the S parameter.  

We move to discussing the solutions of the equations of motion for general $p^2$.
We will obtain the power expansion of the normalizable solution for small $p^2$ and for large Euclidean $p^2$.  
The former can be achieved by separating $K_M(z,p) =  \eta(z) \bar K_M(z,p)$,  
where $\bar K_M$ satisfies the equation   
\beq
\label{e.bkme}
\left (a_M^{-1} \pa_z (a_M \pa_z) + p^2 \right ) \bar K_M(z,p)  = 0  ,
\eeq 
with  the effective warp factor defined as  
\beq
a_M(z) = {\eta^2(z) \over \eta^2(z_0)} a(z) = e^{- 2 \int_{z_0}^z W(z')}.
\eeq 
Since $W(z) > 0$ and asymptotically $W(z) > {\rm const}$, the effective  metric $a_M(z)$ is monotonically decreasing and at least exponentially decaying in the IR, much as the original warp factor $a(z)$.    
We can integrate \eref{bkme} perturbatively in $p^2$, which leads to the following expansion of $K_M$:   
\beq
\label{e.kmsp} 
K_M(z,p) = \eta(z) \left [ 1 
+ p^2 \int_{z_0}^z a_M^{-1} \int_{z'}^\infty a_M 
+ p^4 \int_{z_0}^z a_M^{-1} \int_{z'}^\infty a_M \int_{z_0}^{z''} a_M^{-1} \int_{z'''}^\infty a_M  
+ \co(p^6) \right].
\eeq
In general, this expansion is valid for momenta below the mass gap. 
In order to obtain an expansion for large Euclidean momenta we rewrite    
\beq
K_M(z,i p) = a^{-1/2} e^{- p z} \phi(z) ,
\eeq 
where $\phi$ satisfies the equation 
\beq
\left [e^{2 p z} \pa_z ( e^{-2 p z} \pa_z) - V_M (z) \right ] \phi = 0.   
\eeq 
Again, we integrate perturbatively, this time expanding in powers of $V_M$, 
which yields  
\beq
\label{e.kmlp} 
K_M(z,i p) = a^{-1/2}(z) e^{- p z} \left [ 
1 -  \int_{z_0}^z  e^{2 p z'} \int_{z'}^\infty e^{- 2 p z''} V_M(z'')   + \co(V_M^2)  \right ] .
\eeq
In general, this expansion is valid for $p z_0 \gg 1$, that is for momenta above the UV brane scale.
When the warp factor is approximately AdS near the UV brane, the potential contains a $1/z^2$ term.  
Then the integrals in \eref{kmlp} lead to $\log z$ enhanced terms for $p z_0 < 1$ which undermines the perturbative expansion.    

This ends the math section.  
We are holding all the threads to tackle physics questions. 

\section{Toy Model}
\label{sec:tm} \setcounter{equation}{0} \setcounter{footnote}{0}

We start with a simple toy model based on the $U(1)$ gauge group.
The gauge symmetry is broken both by UV boundary conditions and by a vev of a charged bulk scalar field. 
The spectrum includes a massless Goldstone boson - the gauge-higgs -  that is a mixture of the 5th component of the gauge field and the phase of the bulk scalar. 
The toy model is interesting from the pedagogical point of view 
even though the gauge group is not realistic 
and there is  no dynamics associated with the vev of the gauge-higgs.  
The point is that the lagrangian is simple enough to identify easily all the degrees of freedom.
In particular, the condition for the existence of the gauge-higgs and its equations of motion can be simply derived.   
     
The 5D lagrangian is 
\begin{align}
& \cl =   \sqrt{g}\left\{ -
\frac{1}{4}  X_{MN} X_{MN} 
+ \frac{1}{2}\left|D_M\Phi\right|^2 - V(|\Phi|) 
\right\} ,
\nn 
& X_{MN} =  \pa_M X_N - \pa_N X_M   , 
\nn
& D_M \Phi =   \pa_M \Phi - i g_5 X_M \Phi . 
\end{align}
We parametrize the scalar as 
\beq
\Phi(x,z) = \left [\Lambda(z) +  \phi(x,z) \right ] e^{i g_5 G(x,z)}. 
\eeq
where $\Lambda(z)$ is a $z$-dependent vev that is a solution to the scalar equations of motion. 
The lagrangian becomes 
\begin{align}
\label{e.tml} 
\cl =  & -  \frac{a}{4} X_{\mu\nu} X_{\mu\nu} +  {a \over 2} \left (\pa_z X_\mu - \pa_\mu X_z \right )^2
\nn 
\mbox{} & + \frac{a^3}{2}(\pa_\mu \phi)^2 +   \frac{a^3}{2} g_5^2  (\Lambda + \phi)^2  \left (\pa_\mu G  -  X_\mu \right )^2    
\nn
\mbox{} & - \frac{a^3}{2}(\pa_z \Lambda + \pa_z \phi)^2 -    \frac{a^3}{2} g_5^2  (\Lambda + \phi)^2  \left (\pa_z G  -  X_z \right )^2
\nn
\mbox{} & - a^5 V(\Lambda + \phi).
\end{align}
One can see here that a vev of $X_z$ has no physical significance. Indeed, such a vev must be accompanied by a vev of $G$, with $\la X_z \ra = \partial_z \la G \ra$. 
In the presence of these vevs we can shift, 
\beq
X_z \to \la X_z \ra + X_z = \partial_z \la G \ra + X_z \, ,
\qquad
G \to \la G \ra + G ,
\eeq 
so that they disappear from the lagrangian. In fact, they are a pure-gauge configuration. 

The linear terms in $\phi$ vanish due to the equations of motion for $\Lambda(z)$.  
The quadratic terms are  
\begin{align}
\label{e.tmlq} 
\cl = & - \frac{a}{4} X_{\mu\nu} X_{\mu\nu} +  {a \over 2} \left (\pa_z X_\mu - \pa_\mu X_z \right )^2
\nn 
& \mbox{} + \frac{a^3}{2}(\pa_\mu \phi)^2 +   \frac{a}{2} M^2(z)  \left (X_\mu  - \pa_\mu G \right )^2    
\nn
& \mbox{} - \frac{a^3}{2}(\pa_z \phi)^2 -     \frac{a}{2} M^2(z) \left (\pa_z G  -  X_z \right )^2
\nn
& \mbox{} -  {1 \over 2}a^5 V''(\Lambda) \phi^2.  
\end{align}  
where $M^2(z) = g_5^2 a^2 \Lambda^2$. 
From the above we can see that the 4D effective theory contains the following degrees of freedom  
\bi 
\item U(1) gauge fields $X_{\mu,n}$ living in $X_\mu$. 
\item Scalars and pseudoscalars living in $G$ and $X_z$ that mix with one another:   
\bi 
\item physical pseudo-scalars $P_n$,
\item Goldstones $G_n$ eaten by the massive gauge fields, 
\item depending on the boundary condition for $X_\mu$, physical massless scalar $h$ referred to as the gauge-higgs.
\ei 
\item Scalar fields $\phi_n$ living in $\phi$. 
\ei
Let us discuss them in turn. 

\subsection{Gauge Boson}

The 5D gauge field  can be expanded into the KK modes as,  
\begin{align} 
& X_\mu(x,z) = X_{\mu,n}(x) f_{X,n}(z) ,
\nn
& a^{-1}\pa_z (a \pa_z  f_{X,n}) -  M^2(z) f_{X,n} + m_n^2  f_{X,n}  = 0 ,
\nn 
& a f_{X,n} \pa_z f_{X,n}|_{z = z_0}  = 0  ,
\nn 
& \int_{z_0}^\infty a f_{X,n} f_{X,m} = \delta_{nm} . 
\end{align}
The UV boundary conditions for $f_{X,n}$ can be either Neumann or Dirichlet. 
Here we choose the Dirichlet one, $f_{X,n}|_{z = z_0} = 0$, because it will allow the gauge-higgs to exist.  

The equation of motion for the gauge field profile $f_{X,n}(z)$  has two independent solutions. 
In section 2 we defined $K_M(z,p)$ as the normalizable solution, $\int_{z_0}^\infty a K_M^2 < \infty$. 
Using this notation, the gauge profile can be written as 
\beq
f_{X,n} = \alpha_{X,n} K_M(z,m_n),  
\eeq
and the KK masses are found by solving for the UV boundary condition 
\beq
K_M(z_0,m_n) = 0 .
\eeq

\subsection{Gauge-Higgs}
\label{s.gh}

The 5D fields $G$ and $X_z$ may contain a massless scalar mode - the gauge-higgs - that is embedded as 
\beq
X_z(x,z) \to  h(x)  \pa_z \eta (z) 
\qquad 
G(x,z) \to  h(x)  \eta (z).  
\eeq 
This particular embedding ensures that $h$ does not pick up a mass term from the 
$(\pa_z G  -  X_z)^2$ in the lagrangian \erefn{tml}.   
Furthermore, the mixing term between the gauge-higgs and the vector field reads 
\beq
\cl =   X_\mu \pa_\mu h \left (  \pa_z (a \pa_z \eta)  - a M^2  \eta  - (a \pa_z \eta)|^{\infty}_{z_0} \right )  . 
\eeq 
The UV boundary term vanishes because the gauge field vanishes on the UV brane, while the IR boundary term vanishes for normalizable solutions. 
The gauge-higgs does not mix with the tower of the gauge fields if its profile satisfies the equation  
\beq 
\label{e.gheom}
a^{-1}\pa_z (a \pa_z \eta) - M^2(z) \eta   = 0.  
\eeq
This is the same as the gauge equation of motion with $m_n = 0$, 
and the solutions were discussed  in Section 2 below \eref{ee}. 
Furthermore, the gauge-higgs profile satisfies the normalization condition 
\beq
1 = \int_{z_0}^{\infty} a \left ( (\pa_z \eta)^2 + M^2(z) \eta^2 \right ) ,
\eeq  
which, upon integration by part and using the equation of motion can be written as 
\beq
a(\infty)\eta(\infty)  \pa_z \eta (\infty)  - \eta(z_0)  \pa_z \eta (z_0)  = 1 . 
\eeq  
The first term must vanish for a normalizable solution. 
Consequently, the normalized profile can be written as  
\beq
\eta(z) =  {1 \over \sqrt{U_M(z_0)}}  e^{ - \int_{z_0}^z U_M(z')} ,
\eeq
where $U_M$ is the mass superpotential introduced in \eref{mse}. 
Recall that we proved that $U_M(z) > 0$ in a theory with a mass gap and $M^2 > 0$, 
which implies that the gauge-higgs profile is monotonically decreasing.  
 
Although $\eta(z)$ completely characterizes the gauge Higgs profile, it has no immediate physical meaning. 
Instead, the localization of the gauge-higgs in the gauge field component is determined by $(\pa_z \eta)^2$, 
whereas the localization  in the scalar component is governed by $M^2(z)\eta^2$.   
 
\subsection{Other scalars and pseudoscalars}

We continue the discussion of the scalar spectrum with  the scalar $\phi$, which corresponds to oscillations around the bulk condensate $\Lambda(z)$. $\phi $ does not mix with $G$ or $X_z$ so that it has the simple KK expansion  
$\phi(x,z) = \phi_{n}(x) f_{\phi,n}(z)$.
The profile must solve the equation of motion 
\beq
\left [ a^{-3}\pa_z (a^3 \pa_z) + p^2 - a^2 V''(\Lambda) \right ] f.  
\eeq    
Moreover, diagonalization of the KK action requires vanishing of the boundary term 
\beq
a^3 f_{\phi,n} \pa_z f_{\phi,n}|_{z_0}^\infty  = 0   ,
\eeq 
which leaves us two options: Dirichlet or Neumann boundary conditions on the UV brane (we could obtain mixed boundary conditions if we added UV boundary mass or kinetic terms).  The normalization condition reads  $\int_{z_0}^{\infty} a^3  f_{\phi,n}^2 = 1$. 

The scalar equation of motion is different from that for the gauge field, but similar methods apply. 
We pick up the normalizable solution: $\int^\infty a^3  f^2 < \infty$,  and denote it as $\bar K(z,p)$.
The spectrum is found by imposing the UV boundary condition, e.g. $\pa_z \bar K(z_0,m_n) = 0$. 
Then we can write the profile as $f_{\phi,n} = \bar \alpha_n  \bar K(z,m_n)$.  
We can also rewrite the scalar equation of motion as a Schr\"odinger-type equation by defining  
$\bar f = a^{-3/2}  \bar \Psi$. 
This leads to the equation $( - \pa_z^2 + \bar V) \bar \Psi = p^2 \bar\Psi$ 
with 
$\bar V(z) = a^2 V''(\Lambda) + {3 a''\over 2 a} - {3 (a')^2 \over 4 a^2}$. 
The last two terms are, up to the factor of 3, analogous to the ones in the Schr\"odinger version of the gauge equation of motion.  
As long as  the ''mass term" $a^2 V''$ is positive or vanishing in IR,  the sufficient condition for the scalar spectrum to devolop a mass gap is the same as for the gauge fields: the warp factor should decay as $e^{-\rho z}$ or faster in IR.
 
For the  pseudoscalars living in $X_z$ and $G$, the KK expansion and the equations of motion are more involved. 
We start with the general KK expansion: 
\bea &
X_z(x,z) =  h(x)  \pa_z \eta (z) + G_{n}(x) \ov f_{X,n}(z) + P_n(x)  \ti f_{X,n} ,
\nl 
G(x,z) =  h(x)  \eta (z) +  G_{n}(x) \ov f_{G,n}(z) + P_n(x)  \ti f_{G,n} .
\eea 
The gauge-higgs profile $\eta$ was discussed before. 
The Goldstones $G_n$ should marry the corresponding gauge fields $X_{\mu,n}$, 
such that the quadratic lagrangian depends on the combination $m_n X_{\mu,n} - \pa_\mu G_n$. 
To achieve this, the Goldstone profiles must be synchronized with those of the gauge fields 
\beq
\ov f_{X,n} = m_n^{-1} \pa_z  f_{X,n} \, ,
\qquad
\ov f_{G,n} = m_n^{-1}  f_{X,n}.
\eeq   
Much as the gauge-higgs, the Goldstones cancel out in the $(\pa_z G  -  X_z)^2$ term in \eref{tmlq}, so that they do not pick up any mass terms. 

The rule for the pseudo-scalar profiles is instead that they should {\em not} mix with the gauge fields. 
The mixing terms following from \eref{tmlq} are
\beq
- X_{\mu,m} \pa_\mu P_n \left ( a \pa_z f_{X,m}  \ti f_{X,n} + a M^2 f_{X,m} \ti f_{G,n} \right ) .
\eeq 
So, we need 
\beq
\ti f_{G,n} = {\pa_z (a \ti f_{X,n}) \over a M^2} \, , 
\qquad 
 a f_{X,m}  \ti f_{X,n}| = 0. 
\eeq
Given the above, the kinetic terms and the mass terms are  
\beq
{1 \over 2} \pa_\mu P_n \pa_\mu P_m \ti f_{X,m} a \left [
 \ti f_{X,n} -  \pa_z \left ( {\pa_z (a \ti f_{X,n}) \over a M^2 }  \right )
+   {\pa_z(a \ti f_{X,n}) \over M^2} |  
\right ] , 
\eeq
\beq  
- {1 \over 2} P_n  P_m  a M^2  \left [
\ti f_{X,n} - \pa_z \left ( {\pa_z (a \ti f_{X,n}) \over a M^2 }  \right ) 
\right ]  
\left [
\ti f_{X,m} - \pa_z \left ( { \pa_z (a \ti f_{X,m}) \over a M^2 }  \right ) 
\right ]  .
\eeq 
To diagonalize the kinetic terms we need the orthogonality relation  
\beq
\int_{z_0}^{\infty} a  \ti f_{X,m} \left [ \ti f_{X,n} - \pa_z  \left ( {\pa_z (a \ti f_{X,n}) \over a M^2}\right )  \right ]  = \delta_{nm} 
\eeq 
and the boundary conditions
\beq
M^{-2} \ti f_{X,n} \pa_z (a \ti f_{X,n})|  = 0,   
\eeq 
which leave two options for the UV boundary conditions. 
The mass terms are diagonalized and the orthogonality relations are fulfilled if the profile $\ti f_{X,n}$ is a solution of the equation 
\beq
\left [ M^2(z) \pa_z { 1 \over a M^2(z)} \pa_z a  + p^2 -  M^2(z)  \right ] \ti f  = 0. 
\eeq
We apply the same methods all over again.  
We pick up the normalizable solution: $\int^\infty a \ti f^2 < \infty$,  and denote it as $\ti K_M(z,p)$.
The spectrum is found by imposing the UV boundary condition, e.g. $\ti K_M(z_0,m_n) = 0$. 
Then we can write the profile as $\ti f_{X,n} = \ti \alpha_n \ti K_M(z,m_n)$.  
We can also rewrite the pseudoscalar equation by defining 
$\ti f = a^{-1/2} M\ti \psi$
which leads to the  Schr\"odinger  equation   with $\ti V = M^2 + a^{1/2} M \pa_z^2 (a^{-1/2} M^{-1})$.
For a non-pathological behavior of $M^2$ in IR, the exponential decay of the warp factor in IR ensures the presence of a mass gap in the pseudoscalar spectrum. 
  

\section{Soft-Wall Model of Electroweak Breaking}
\label{sec:eb} \setcounter{equation}{0} \setcounter{footnote}{0}

In this section we discuss a model that accommodates the SM electroweak gauge bosons and the Higgs sector.
Many features of the toy model in Section 3 carry over to the realistic setting.
In particular, the profile of the gauge-higgs is unchanged.  
The main complication is that the vev of the gauge-higgs affects the mass spectrum and the KK decomposition.

The simplest model that includes the electroweak group and custodial symmetry is based on $SO(5)$ gauge symmetry \cite{ACP}.  
We consider here the $SO(5) \times U(1)_X$ gauge theory, where  $SO(5)$ is broken down to $SO(4)$ by a vev of a real bulk scalar transforming as $\bf 5_0$. 
The UV boundary conditions break $SO(5) \times U(1)_X$ down to $SU(2)_L \times U(1)_Y$, where the electroweak group is a subgroup of $SO(4) \times U(1)_X$ left unbroken by the bulk scalar vev. 
The lagrangian is given by  
\begin{align}
& \cl =  \sqrt{g}\left\{ 
-\frac{1}{4}  {\tr} A_{MN} A_{MN}  
+ \frac{1}{2} D_M\Phi^T D_M\Phi  - V(\Phi) 
\right\},
\nn 
& A_{MN} =  \pa_M A_N - \pa_N A_M - i g_5 [A_M,A_N] , \qquad A_M = A_m^\alpha T^\alpha, 
\nn 
& D_M \Phi =   \pa_M \Phi - i g_5 A_M  \Phi .  
\end{align}
The  $SO(5) \times U(1)_X$ generators 
are normalized as $\tr T^\alpha T^\beta = \delta^{\alpha\beta}$. 
We split these generators into four classes: 
$T_L^a$ and $T_R^a$, $a = 1 \dots 3$ that generate the  $SO(4) \equiv SU(2)_L \times SU(2)_R$ 
subgroup of $SO(5)$, 
$T_C^\ha$, $\ha = 1 \dots 4$ that generate the $SO(5)/SO(4)$ coset, 
and $T_X \equiv I$ for  $U(1)_X$.
The $SO(5) \times U(1)_X$ gauge field can be analogously split into $L_M^a$, $R_M^a$, $C_M^\ha$, $X_M$.  
The bulk scalar is parametrized as 
\beq
\Phi(x,z) = \left [\Lambda(z) +  \phi(x,z) \right ] e^{i g_5 G_\ha(x,z) T_C^\ha} \bvec \vec 0 \\ 1  \evec  
=  \left [\Lambda +  \phi \right ] \bvec
{G_a \over G} \sin (g_5 G/\sqrt{2})  \\
{G_4 \over G} \sin (g_5 G/\sqrt{2})  \\
\cos (g_5 G/\sqrt{2})  
    \evec  ,
\eeq 
where $G^2 = G_\ha G_\ha$. The vev $\Lambda(z)$  breaks $SO(5)$ to $SO(4)$ and gives the mass $M^2 = a^2 \Lambda^2 g_5^2/2$ to the coset gauge bosons $C_M^\ha$. 

For the time being, we do not specify the UV boundary conditions for the bulk scalar: they can be Dirichlet, or Neumann, or mixed. 
Since the gauge-higgs live partly in the bulk scalar, one may want to impose the Dirichlet boundary condition $\Phi(z_0) = 0$. 
This would protect us from  UV brane localized $SO(5)$ violating mass terms for $\Phi$ that would imply mass terms for our gauge-higgs.\footnote{Thanks to Csaba Csaki for pointing this out.}   
For our purpose, however, it is sufficient if the scalar vev is peaked in IR while it is Planck suppressed on the UV brane.  This will ensure that the gauge-higgs component in $\Phi(z_0)$ is small enough so that the hierarchy problem is not reintroduced.  

We fix the vev of the fifth component of the gauge field to be along the $SO(5)$ generator $T_C^4$.  
Much like in the toy model, there is a physical scalar mode embedded in $C_z^4$ and $G_4$:     
\beq
X_z(x,z) \to  h(x)  \pa_z \eta (z) ,
\qquad 
G(x,z) \to  h(x)  \eta (z).  
\eeq 
where $\eta(z)$ is the gauge-higgs profile discussed at length in Section \ref{s.gh}. 
We identify this scalar mode with the SM higgs boson. 
We assume that it acquires a vev $\la h(x) \ra = \ti v$. 
In principle, this vev is not a free parameter but it is fixed by the parameters of the 5D model, as it is obtained by minimizing the radiative Coleman-Weinberg potential for $h$.
Typically, the largest contribution to the potential comes from the loops of the top quark and its KK modes, and the vev depends mostly on the parameters in the top sector. 
In this paper we do not study the fermionic sector nor the one-loop dynamics, postponing the complete discussion to future publications.        

\subsection{Vector Spectrum}

Unlike in the toy model, the gauge-higgs vev affects the spectrum of the theory, 
in particular, it gives masses to some of the zero mode gauge bosons.  
The quadratic lagrangian for the gauge fields in the gauge-higgs background reads 
\begin{align}
\label{e.ql}
\cl = & 
-\frac{a}{4}  {\tr} A_{\mu\nu} A_{\mu\nu}   - {a \over 2} \tr D_z A_\mu D_z A_\mu
\nn
& \mbox{} +  {a^3 \over 2} M^2  (C_\mu^{4})^2  
+ {a^3 \over 2}  M^2  \left ( C_\mu^a \cos_z   +   {1 \over \sqrt{2}}  (L_\mu^a - R_\mu^a) \sin_z \right )^2   .
\end{align}
\bea
D_z L_\mu^a  &= & \pa_z L_{\mu}^a -  {g_5 \ti v \over 2} C_{\mu}^a \pa_{z} \eta ,
\nn 
D_z R_\mu^a  &= & \pa_z R_{\mu}^a + {g_5 \ti v \over 2} C_{\mu}^a \pa_{z} \eta ,
\nn 
D_z C_\mu^a  &= & \pa_z C_{\mu}^a  +  {g_5 \ti v\over 2} (L_{\mu}^a - R_{\mu}^a) \pa_{z} \eta ,
\nn 
D_z C_\mu^4  & = &  \pa_z C_{\mu}^4 , \qquad  D_z X_\mu  =  \pa_z X_{\mu},
\nn
\sin_z & \equiv & \sin (g_5 \ti v \eta(z)/\sqrt{2}) .
\eea  
One can see that $\ti v$ mixes $L_\mu^a$, $R_\mu^a$, $C_\mu^a$ with each other.  
Thus, the mass eigenstates in the presence of the vev will be embedded in all these fields (and also in $X_\mu$, which mixes with the others via UV boundary conditions).   
Therefore, we write the KK expansion as  
\bea 
L_\mu^a(x,z) &=& A_{\mu,n}(x) f_{L,n}^a(z) ,
\nn
R_\mu^a(x,z) &=& A_{\mu,n}(x) f_{R,n}^a(z) ,
\nn
C_\mu^\ha(x,z) &=& A_{\mu,n}(x) f_{C,n}^\ha(z) , 
\nn
X_\mu(x,z) &=& A_{\mu,n}(x) f_{X,n}(z) .
\eea 
The profiles satisfy the UV boundary conditions, that reduce $SO(5) \times U(1)_X$ down to $SU(2)_L \times U(1)_Y$:  
\beq
\pa_z f_L^a(z_0)  = 0 , \qquad
 f_R^i(z_0)  = 0 , \qquad  f_C^\ha(z_0)  = 0 ,
\eeq 
\beq
c_x f_R^3(z_0) - s_x f_X(z_0) = 0 , \qquad 
s_x \pa_z f_R^3(z_0) + c_x \pa_z f_X(z_0) = 0 ,  
\eeq   
where $s_x = g_X/\sqrt{g_X^2 + g_5^2}$, $c_x = g_5/\sqrt{g_X^2 + g_5^2}$.
We can identify $s_x^2 = g_Y^2/g_L^2$.   
The kinetic terms  fix the normalization condition: 
\beq
1 = \int_{z_0}^{\infty} dz  a(z) \sum_{A=L,R,C,X} (f_A(z))^2  .
\eeq  

The equations of motion are complicated because the various SO(5) gauge bosons are mixed by the vev of the gauge-higgs. 
The usual trick employed in the gauge-higgs models is to simplify the equations of motions by an appropriate rotation in the group space. 
In the present case, the rotation  is given by 
\bea
\label{e.wr}
f_{L}^a &=& {1 + \cos_z \over 2} \hat f_{L}^a +  {1 - \cos_z \over 2} \hat f_{R}^a + {\sin_z \over \sqrt 2} \hat f_C^a ,
\nn 
f_{R}^a &=& {1 - \cos_z \over 2} \hat f_{L}^a +  {1 + \cos_z \over 2} \hat f_{R}^a - {\sin_z \over \sqrt 2} \hat f_C^a ,
\nn
f_{C}^a &=&  - {\sin_z \over \sqrt 2} \hat f_L^a + {\sin_z \over \sqrt 2} \hat f_R^a + \cos_z \hat f_C^a ,
\nn
f_{C}^4 &=& \hat f_C^4 , \qquad f_{X} = \hat f_X . 
\eea
This transformation {\em locally} removes the gauge-higgs vev from the lagrangian (so that it affects the spectrum only {\em non-locally}, via the boundary conditions).
As a consequence, the hatted profiles satisfy equations of motion that do not depend on $\ti v$,  
\bea  
\label{e.hem}
& \left (a^{-1} \pa_z (a \pa_z) + m_n^2 \right ) \hat f_{L,R,X} = 0 ,
\nl 
\left (a^{-1} \pa_z (a \pa_z) + m_n^2 - M^2 \right ) \hat f_{C} = 0 .
\eea 
Note that $\eta(z)$ vanishes in the IR, thus $f = \hat f$ asymptotically for $z \to \infty$.
Therefore, we should pick the solution to the equations of motion \erefn{hem} that decays in IR. 
We write the hatted profiles $\hat f_{L,R,X} = \alpha_{L,R,X} K_0(z,m_n)$, $\hat f_{C} = \alpha_{C} K_M(z,m_n)$. 
From that, the true profiles $f$ can  be obtained through \eref{wr}.  
The constants $\alpha$ are determined up to normalization by the UV boundary conditions.   
It is clear that they depend on the gauge-higgs vev only via  $\sin (g_5 \ti v \eta(z_0)/\sqrt{2})$.
Writing it as  $\sin (\ti v/f_h)$ defines the global symmetry breaking scale:  
\beq
\label{e.f}
f_h = {\sqrt{2} \over g_5 \eta(z_0)} 
=  {\sqrt{2 U_M(z_0)} \over g_5} .
\eeq
When $\sin (\ti v/f_h) \ll 1$, the gauge-higgs becomes  SM-like.  
The other extreme limit is \mbox{$\sin(\ti v/f_h) = 1$}, in which the electroweak sector is Higgsless for all practical purposes (even though a light scalar particle exists, it does not couple linearly to W and Z bosons, so it cannot unitarize the longitudinal gauge boson scattering). 
In between these two extremes is the pseudo-Goldstone Higgs, where the Higgs boson partly (up to $E^2/f_h^2$ terms) unitarizes the longitudinal scattering.  
  
The UV boundary conditions relate the constants $\alpha$ and yield the quantization condition on the mass $m_n$. 
There are several classes of solutions to the UV boundary conditions, that define the towers of KK modes. 
Below we write down only those that contain a massless mode in the limit of $\ti v \to 0$. 
The lowest solutions of the quantization condition are identified with the SM electroweak gauge bosons.   

\bi 
\item  {\bf Photon tower}. \\   
This is a tower of neutral gauge bosons where the hatted profiles are given by 
\bea
\hat f_L^3 &=& {s_x \over \sqrt {1 + s_x^2} } \alpha_{\gamma} K_0(z,m) ,
\nn
\hat  f_R^3 &=& {s_x \over \sqrt {1 + s_x^2} }  \alpha_{\gamma}  K_0(z,m) , 
\nn
f_X &=&   {c_x \over \sqrt {1 + s_x^2} }  \alpha_{\gamma} K_0(z,m) ,
\eea 
while the remaining profiles all vanish.  
$\alpha_\gamma$  is given by the normalization condition 
$(\alpha_{\gamma})^{-2}  =   \int_{z_0}^\infty a(z)  K_0(z,m)^2$.   
The spectrum of the photon and its KK modes is given by the quantization condition 
\beq
 K_0'(z_0,m) = 0 .
\eeq
$m = 0$ is always a solution to the above, and the corresponding mode is identified with the SM photon. 
In that case $K_0(z,0) = 1$ and  $\alpha_\gamma = L^{-1/2}$.

\item {\bf $W$ tower}. \\  
Another solution is a tower of charged gauge bosons. 
The profiles are given by 
\bea
\hat f_L^i &=& \alpha_W {1 + \cos(\ti v/f_h) \over 2} K_0(z,m) , 
\nn
\hat  f_R^i &=& \alpha_W {1 - \cos(\ti v/f_h) \over 2} K_0(z,m) ,
\nn
\hat  f_C^i &=& \alpha_W {\sin(\ti v/f_h) \over \sqrt{2}} {K_0(z_0,m) \over K_M(z_0,m)} K_M(z,m) . 
\eea 
The quantization condition  depends on $\ti v$,  
\beq
\label{e.wqc} 
 {K_0'(z_0,m) \over K_0(z_0,m)} =  
   {\sin^2(\ti v/f_h) \over 2} \left (  - {K_M'(z_0,m) \over K_M(z_0,m)} + {K_0'(z_0,m) \over K_0(z_0,m)} \right ) .
\eeq
In the limit $\ti v \to 0$, there is a massless mode. 
In the presence of electroweak breaking the lowest solution  becomes $m_W \approx {g_L f_h \over 2} \sin (\ti v/ f_h)$ and the corresponding mode is identified with the SM W boson.


\item {\bf $Z$ tower}. \\  

This is a tower of neutral gauge bosons where, unlike for the photon tower, the masses depend on the gauge-higgs vev.  
The profiles are given by
\bea
\hat  f_L^3 &=& \alpha_Z {c_x^2 + (1+ s_x^2)\cos(\ti v/f_h) \over 2 (1 + s_x^2)^{1/2}} K_0(z,m) ,
\nn
\hat f_R^3 &=& \alpha_Z {c_x^2 - (1+ s_x^2)\cos(\ti v/f_h) \over  2 (1 + s_x^2)^{1/2} } K_0(z,m) ,
\nn
\hat  f_C^3 &=& \alpha_Z 
 {\sin(\ti v/f_h) \over  \sqrt{2}}  (1 + s_x^2)^{1/2} {K_0(z_0,m) \over K_M(z_0,m)} K_M(z,m) ,
\nn
f_X &=&  - \alpha_Z {s_x c_x \over  (1 + s_x^2)^{1/2} }  K_0(z,m) . 
\eea 
The quantization condition reads
\beq
{K_0'(z_0,m) \over K_0(z_0,m)} 
 =   {(1 + s_x^2) \sin^2(\ti v/f_h) \over 2} \left (  - {K_M'(z_0,m) \over K_M(z_0,m)} + {K_0'(z_0,m) \over K_0(z_0,m)} \right ) .
\eeq
The lowest lying solution is identified with the SM Z boson. 
\ei 
There are other classes of solutions to the UV boundary conditions that do not lead to  zero modes, and where the mass spectrum is insensitive to the gauge-higgs vev.   
Furthermore, similarly as in the toy model there are scalar, pseudo-scalar and unphysical Goldstone particles. 
These are however not important for the following and we omit this discussion.   

\subsection{Gauge Contributions to the Higgs Mass}

We turn to discussing the one-loop corrections to the Higgs mass generated by the electroweak gauge bosons and its KK modes.
Our gauge-Higgs is a (pseudo-) Goldstone boson of SO(5) symmetry broken spontaneously down to SO(4) in the bulk, and explicitly down to SU(2) on the UV brane.     
As a consequence of that, the Higgs potential vanishes at the tree-level as long as there is no SO(5) breaking operator involving the scalar $\Phi$ localized on the  UV brane. 
A radiative Higgs potential is generated at one-loop level  because the tree-level KK mode masses depend on the vev $\ti v$.
The simplest way to calculate it is to use the spectral function $\rho(p^2) = \det (-p^2 + m_n^2(\ti v))$, which is
a function of 4D momenta with zeros encoding the whole KK spectrum in the presence of electroweak  breaking \cite{AA}. 
With  a spectral function at hand, we can compute the Higgs potential from the Coleman-Weinberg formula, 
\beq
\label{e.cws0} 
V(\ti v) = {N \over (4 \pi)^{2}}  \int_0^\infty dp \, p^{3} \log \left ( \rho[-p^2] \right ) ,
\eeq
where $N = + 3$ for gauge bosons.

From the quantization conditions obtained in the previous section we know that the spectral functions for 
W and Z towers have the form $\rho_{W,Z} = 1 + F_{W,Z}(p^2) \sin^2(\ti v/f_h)$.
It is convenient to define the form factor 
\beq
\label{e.ff}
\Pi_M(p^2) = \pa_z \log \left (K_M(z,p) \right)|_{z = z_0} .
\eeq 
From \eref{wqc} we identify 
\beq
\label{e.gff}
F_W(p^2)  =  {\Pi_M(p^2) - \Pi_0(p^2)  \over 2 \Pi_0(p^2)} 
\eeq 
and $F_Z(p^2) = (1 + s_x^2) F_W(p^2)$. 
$F_W$ determines the one-loop gauge contribution to the Higgs mass parameter:  
\beq
m_H^2 \equiv {\pa^2 V \over \pa \ti v^2}|_{\ti v= 0} = 
{3(3+ s_x^2) \over f_h^2 (4 \pi)^{2}}  \int_0^\infty dp \, p^{3} F_W(-p^2) 
\eeq   
Quadratic divergences are avoided if the form factor $F_W$ decays faster than $1/p^2$ for large Euclidean momenta.
At this point our efforts from section 2 are beginning to pay off.
At small Euclidean momenta we can use \eref{kmsp} to find that $F_W(-p^2) \approx g_5^2 f_h^2/4 L p^2$.
This can be identified with the SM W and Z boson contribution to the Higgs mass.  
The presence of the tower of KK modes changes the shape of $F_W(-p^2)$ for momenta above the mass gap. 
Whether the tower cuts off the quadratic divergences depends on the asymptotic form of  $F_W(-p^2)$  for $-p^2 \to \infty$.       
Using \eref{kmlp} we find that the leading asymptotic behavior of the form factor is given by
\beq
F_W(-p^2) \approx  { e^{2 p z_0} \int_{z_0}^\infty e^{- 2 p z} M^2 (z) \over 2 p  + a'/a  + \dots} .
\eeq   
The integral is exponentially suppressed in the IR, so that we can expand the mass term around the UV brane as 
$M^2(z) = M^2(z_0) + (z-z_0)\pa_z M^2(z_0) + \dots$ and integrate term by term.  
If $M^2(z_0) \neq 0$, the leading asymptotic behavior of the warp factor is 
\beq
F_W(-p^2)  \approx  {M^2 (z_0) \over 4 p^2} + \co(1/p^3) ,  
\qquad 
p z_0 \gg 1 .
\eeq 
Thus the one-loop Higgs mass is in general quadratically divergent, $\delta m_H^2 
\sim M^2 (z_0) \Lambda^2 /16\pi^2 f_h^2 \sim M^2(z_0)/ (z_0 f_h)^2$, unlike in the standard hard-wall gauge-higgs scenario.
We can avoid quadratic divergences if $M^2(z_0) = 0$,
which is automatic when  the bulk scalar that breaks SO(5) satisfies Dirichlet boundary conditions on the UV brane.     
More precisely, the mass is finite when we impose the condition that $M^2$ vanishes faster than $(z - z_0)^2$ in the vicinity of the UV brane, 
while  for $M^2(z) \sim (z- z_0)^2$ the Higgs mass is logarithmically sensitive to the cutoff. 
In practice, we do not need to insist on the finiteness of the Higgs mass: 
it is enough if $M^2(z_0)$ is sufficiently suppressed, $M^2(z_0) \sim \co(z_0^2 \tev^4)$.  
We note in passing that the UV behaviour of $M^2(z)$ is related to the dimension of the operator breaking the $SO(5)$ symmetry in the holographic dual. The latter should be larger than $2$ to avoid quadratic divergences.   
        

\subsection{Oblique Corrections}

Integrating out the heavy KK modes leaves the SM lagrangian plus higher-dimension operators.
We will assume here that all the light SM fermions are localized on the UV brane. 
In such a case the corrections to the SM lagrangian are universal (except the operators involving the third generation fermions), which means that there exists a field redefinition such that the higher dimension operators involve only the electroweak gauge boson and the Higgs field, whereas vertex correction and four-fermion operators are absent.  
Phenomenologically, the most important are the corrections to the quadratic terms involving the $SU(2)_L \times U(1)_Y$ gauge bosons
(there are also corrections to the triple and quartic gauge boson vertices, but these are less constrained by experiment).
Restricting to the lagrangian terms with at most four derivatives, these corrections can be described by seven "oblique" parameters \cite{BPRS}. 
We define them as follows: 
\begin{align}
\label{e.p4} 
\cl_{eff} = &  
-  {1 \over 4} (L_{\mu \nu}^a)^2   -  {1 \over 4} (B_{\mu \nu})^2  
+ {g_L^2 v^2 \over 8} L_\mu^i L_\mu^i    
\nn
& \mbox{} +   {v^2 \over 8} \left ( g_L L_\mu^3  - g_Y B_\mu \right ) 
\left ( 1 -   \alpha_T {g_Y^2 v^2  \over 2} + \alpha_U \pa^2  + \alpha_V \pa^4\right )  \left ( g_L L_\mu^3  - g_Y B_\mu \right ) 
\nn 
& \mbox{} -    {1 \over 4} \alpha_W (\pa_\rho L_{\mu\nu}^a)^2  -   {1 \over 4} \alpha_Y (\pa_\rho B_{\mu\nu})^2 
-  {g_L g_Y v^2 \over 8} L_{\mu\nu}^3 \left( \alpha_S + \alpha_X \pa^2 \right ) B_{\mu\nu}
+ \co (\pa^6) .
\end{align}
As explained in \cite{BPRS}, the parameters $\alpha_{T,S,W,Y}$ are most relevant  for phenomenologists, since they are the lowest order in their class (they also correspond to dimension six operators in the effective SM lagrangian).   
Furthermore, in our set-up $\alpha_T = 0$. 
This a consequence of the original SO(5)/SO(4) coset structure, which implies that the quadratic terms in the effective lagrangian respect the SU(2) custodial symmetry rotating the triplet $L_\mu^a$. 
This leaves us with three oblique parameters which we  will find by matching with the low-energy effective action obtained by integrating out the KK modes.    
The oblique parameters of ref. \cite{BPRS} are related to the dimensionful coefficients $\alpha$ by 
$\hat {S} = m_W^2 \alpha_S$, 
$\hat T =  {g_Y^2 v^2 \over 2} \alpha_T$,  
$W =    m_W^2 \alpha_W$,  
$Y =    m_W^2 \alpha_Y$. 
The Peskin-Takeuchi S parameter is $S =  4 \pi v^2  \alpha_S$.   


The derivation of the low-energy effective action using the ''holographic approach" \cite{BPR,ACP,PW} is shifted to Appendix \aref{hd}. 
It turns out that the quadratic part of the low-energy effective action can be expressed in terms of the form factor defined in \eref{ff},  
\begin{align}
\label{e.gel} 
\cl_{eff} = &  - {1 \over 2} \Pi_0(p^2) \left [ 
Z_{L}^{-1} L_\mu^a L_\mu^a   + Z_{B}^{-1} B_\mu B_\mu \right ] 
\nn
& \mbox{} + Z_{L}^{-1} {\sin^2(\ti v/f_h) \over 4} \left (\Pi_0(p^2) -\Pi_M(p^2) \right )
\left [ 
L_\mu^i L_\mu^i + (L_\mu^3 - {g_Y \over g_L} B_\mu )^2 \right ] 
\end{align}
where $Z_{L,B}$ are arbitrary normalization factors.   
In the limit of no EW breaking, $\ti v = 0$, the gauge bosons should be massless, which is true when $\Pi_0(0) = 0$.
This is a consequence of the 5D equations of motion, namely that the gauge invariance is left unbroken for $M^2 = 0$.  
Note also that, from \eref{f}, the global symmetry breaking scale $f_h$ can be expressed by the form factors as 
\beq
f_h^2 =  - {2 \Pi_M(0) \over g_5^2} =   - {2 \Pi_M(0) \over Z_L g_L^2}  .
\eeq 

Expanding the form factors in powers of $p^2$ we match the above lagrangian with \eref{gel}. 
Canonical normalization is achieved when the normalization factors are chosen as 
\bea
\label{e.zl}
Z_L &=& \Pi_0'(0) + {\sin^2(\ti v/f_h) \over 2} \left (\Pi_M'(0) -\Pi_0'(0) \right) ,
\nn
Z_B &=& \Pi_0'(0) {\Pi_0'(0) + {\sin^2(\ti v/f_h) \over 2} \left (\Pi_M'(0) -\Pi_0'(0) \right) \over 
\Pi_0'(0) + {\sin^2(\ti v/f_h) \over 2} (1 - g_Y^2/g_L^2) \left (\Pi_M'(0) -\Pi_0'(0) \right)} .
\eea 
Then the W mass is given by $m_W^2 = - Z_L^{-1} \Pi_M(0) \sin^2(\ti v/f_h)/2$ which allows us to identify the electroweak scale as 
\beq
v^2 = -{2 \Pi_M(0) \over Z_L g_L^2} \sin^2(\ti v/f_h)  = f_h^2 \sin^2 (\ti v/f_h) .
\eeq 
The W and Z masses are positive if $\Pi_M(0) < 0$ which will turn out to be a general consequence of the 5D equations of motion. 
We can see that $v \approx \ti v$ for $\sin (\ti v/f_h)\ll 1$, while in the technicolor limit, $\sin(\ti v/f_h) = 1$, we obtain $v = f_h$. 
As we mentioned,  $v/f_h$ is determined by the radiatively generated Higgs potential, but we don't discuss this issue in this paper.  

Further expanding the form factors we read off the oblique parameters defined in \eref{p4}:
\beq
\alpha_S  = {\Pi_M'(0) - \Pi_0'(0) \over \Pi_M(0)} ,
\qquad 
\alpha_W  =  {\Pi_0''(0) \over 2 Z_L} ,
\qquad 
\alpha_Y  =  {\Pi_0''(0) \over 2 Z_B} .
\eeq  
If the solutions to the 5D equations of motion can be found then we can write down the full form factor $\Pi_M(p^2)$ using \eref{ff} and trivially compute the oblique parameters.   
In those (more typical) instances when the explicit solution  is not known we can still learn a great deal by using the results from section 2. 
In particular, the small momentum $p^2$ given in eq. \eref{kmsp} implies the following ''chiral expansion" of the form factors:  
\begin{align}
& \Pi_0(p^2) = p^2 L \left ( 1  
+  p^2 {\int_{z_0}^\infty a \int_{z_0}^{z} a^{-1} \int_{z'}^\infty a \over  \int_{z_0}^\infty a }
\right ) + \co(p^6) ,
\\
&
\Pi_M(p^2) =   
{\eta'(z_0) \over \eta(z_0)} +   p^2 \int_{z_0}^\infty a(z) {\eta^2(z) \over  \eta^2(z_0)} + \co(p^4) .
\end{align}
$\Pi_0(0) = 0$, as promised.  
The first derivative is set by the invariant length of the 5D, $\Pi_0'(0) = \int_{z_0}^\infty a(z) = L$, while the second derivative is also positive and is given by more complicated functional of the warp factor.
It follows that the oblique parameters $W$ and $Y$ are always positive in the 5D set-up.  
In the limit $\sin(\ti v/f) \to 0$ they are equal and given by
\beq
\alpha_W = \alpha_Y =  {\int_{z_0}^\infty a \int_{z_0}^{z} a^{-1} \int_{z'}^\infty a \over  \int_{z_0}^\infty a} .
\eeq  
$\Pi_M(p^2)$  depends on the SO(5) breaking bulk dynamics personified in the gauge-higgs profile $\eta(z)$.
We have proved earlier that $\eta(z)$ is monotonically decreasing everywhere in the bulk, as there is a mass gap and $M^2 > 0$ everywhere. 
In particular, $\Pi_M(0) < 0$, which is what it takes to ensure $m_W^2 >0$. 
Furthermore, the perturbative expression for $\Pi_M(p^2)$ leads to the "dispersive" representation of the S parameter 
\beq
\label{e.asd} 
\alpha_S  = \int_{z_0}^\infty a(z) \left ( \eta^2(z_0) -  \eta^2(z) \right )    
\qquad 
\eeq
The decrease of $\eta(z)$ implies  that {\em the S parameter in the 5D set-up is always positive}.
The no-go theorem for negative S  was proved in ref. \cite{BPR} for Higgsless models with a RS background, and then in ref. \cite{ACGR} in the context of 5D models with the $SU(2)_L \times SU(2)_R \times U(1)_X$ bulk gauge symmetry, for the Higgs field realized as an IR brane or a bulk scalar, and for an arbitrary warped metric.  
Our analysis extends these results to the 5D soft wall models of gauge-higgs unification, where the Higgs is a pseudo-Goldstone boson living partly in a bulk scalar and partly in a fifth component of a gauge field.  
Eq.\erefn{asd} and the positivity of S hold for arbitrary $v/f_h$, in particular in the Higgsless limit $v = f_h$. 
Therefore, the soft wall does not allows us to evade the no-go theorem for negative S.

\section{Examples}
\label{sec:na} \setcounter{equation}{0} \setcounter{footnote}{0}

\subsection{Hard Wall}

As a reference point, we review the results for the standard hard wall RS set-up, where the 5D coordinate is cut off at finite $z = z_L$ by the IR brane, and it is also the IR brane rather than  a bulk scalar that breaks $SO(5) \to SO(4)$.
The hard wall can be viewed as a special limit of the soft wall: the warp factor is discontinuous and vanishes for $z\geq z_L$, 
while the symmetry breaking mass $M^2$ is zero everywhere except for $z = z_L$ where it is infinite.  
Thus, our soft wall analysis applies to the hard wall as well after some obvious adjustments.
Namely, the solutions $K_M(z)$ and $K_0(z)$ should be chosen such as to satisfy the appropriate IR boundary (rather than normalizability) conditions.
Breaking of the global symmetry by the IR brane amounts to setting $K_M(z_L,p)=0$. 
For $K_0$ we impose the mixed boundary conditions $a(z_L) K_0'(z_L,p) = p^2 r K_0(z_L,p)$, 
where $r$ is as an IR brane kinetic term (common to SO(4) and $U(1)_X$, for simplicity).   
These boundary conditions  imply that $K$'s can be expanded in powers of $p^2$ as 
\bea
K_0(z,p) &=& C_0 \left (  
1  - p^2 \int_{z}^{z_L} a^{-1} \int_{z'}^{z_L} a - p^2 r  \int_{z}^{z_L} a^{-1}  + \co(p^4)
\right) ,
\nn
K_M(z,p) &=& C_0  \left ( 
\int_{z}^{z_L} a^{-1} + p^2 \int_{z}^{z_L} a^{-1} \int_{z_0}^{z'} a \int_{z''}^{z_L} a^{-1}  
+ \co(p^4) \right ) .
\eea   
Everything else follows from that expansion, according to formulas presented in the previous section. 
The form factors are 
\bea
\Pi_0(p^2) &=& p^2(L+r)  + \co(p^4) ,
\nn
\Pi_M(p^2) &=&  -  {1 \over \int_{z_0}^{z_L} a^{-1}}  
+ p^2 {\int_{z_0}^{z_L} a^{-1} \int_{z_0}^{z} a \int_{z'}^{z_L} a^{-1} \over  \left (\int_{z_0}^{z_L} a^{-1} \right )^2}   
+ \co(p^4) .
\eea
The expression for the global symmetry breaking scale follows:   
\beq
f_h^2 = {2 \over g_*^2 z_0 \int_{z_0}^{z_L} a^{-1}} , 
\eeq
where $g_*^2 = g_5^2/z_0$ is the dimensionless bulk gauge coupling. 
In the absence of UV brane kinetic terms the weak coupling is related to $g_*$ via 
\beq
\label{e.wgc}
g_L^2 = g_*^2 {z_0 \over L + r - {\eps^2 \over 2} \left( L + r -   {\int_{z_0}^{z_L} a^{-1} \int_{z_0}^{z} a \int_{z'}^{z_L} a^{-1} \over  \left (\int_{z_0}^{z_L} a^{-1} \right )^2}  \right ) } ,
\eeq   
where $\eps = \sin (\ti v/f_h)$. 
The normalized Higgs profile is given by  
\beq
\eta(z) = {\int_{z}^{z_L} a^{-1} \over \sqrt{\int_{z_0}^{z_L} a^{-1}}} .
\eeq 
The S parameter becomes 
\beq
S = 4 \pi v^2 \left [ 
{(L + r)(\int_{z_0}^{z_L} a^{-1})^2
- \int_{z_0}^{z_L} a^{-1} \int_{z_0}^{z} a \int_{z'}^{z_L} a^{-1} \over \int_{z_0}^{z_L} a^{-1} }
\right] .
\eeq 
For AdS the warp factor is $a = z_0/z$. It follows that $f_h^2 \approx 4/g_{*}^2 z_L^{2}$.    
We obtain the S parameter 
\beq
\label{e.hws} 
S \approx {3 \pi \over 2} v^2 z_L^2 \left ( 1  + {4 r \over 3 z_L} \right )  
\approx \eps^2  {6 \pi \over g_*^2} \left ( 1  + {4 r \over 3 z_L} \right ) ,  
\eeq  
in agreement with ref. \cite{ACP}. 
The first equality shows that, for small IR brane term, the S parameter is of order $v^2/\mkk^2$ where $\mkk$ is the mass of the first KK mode: $\mkk \approx 2.4/z_L$. 
Imposing the constraint $S < 0.2$ leads to the bound on the KK mass $\mkk \simgt 3 \tev$.
Adding a positive IR brane kinetic term makes the bound even stronger. 
The second equality shows that for a fixed $\eps = v/f_h$ the size of the S parameter is determined by the strength of the bulk gauge coupling. 
The latter is related to the weak gauge coupling as in \eref{wgc} but we can control this relation by adjusting $z_0$ or adding a UV gauge kinetic term. 
Perturbativity of the 5D theory constrains $g_*$ such that $N_{CFT} \equiv 16 \pi^2/g_{*}^2 \gg 1$, 
which then implies that $\eps$ has to be small enough.  
For example, for $N_{CFT} = 10$ the bound is $\eps < 0.4$.     
This leads  to some tension with naturalness, since the fine-tuning involved in preparing a  correct  electroweak breaking vacuum is of order $\eps^2$. 

The W and Y parameters are given by (for $r = 0$)
\beq
W \approx Y  \approx {m_W^2 z_L^2 \over 4 \log (z_L/z_0)} .
\eeq 
As long as $z_L/z_0$ is large, which is the case when the set-up addresses the Planck-TeV hierarchy, the log in the denominator is large and $W$ and $Y$ are suppressed with respect to $m_W^2/\mkk^2$.   
Thus, the constraints from $W$ and $Y$ are much weaker than those from $S$. 
For example, for $z_L/z_0 \sim 10^{16}$, imposing $W < 10^{-3}$ yields $\mkk > 500 \gev$.

In the following we investigate if the constraint from the S parameter may be improved in the soft-wall backgrounds.  

\subsection{Linear Soft Wall}

Our first example of electroweak breaking on the soft wall has the metric that yields a linear trajectory for KK resonances.
The warp factor  is \cite{EKSS}
\beq
a(z) = {z_0 \over z} e^{-\rho^2(z^2-z_0^2)}. 
\eeq
We assume $1/z_0 \gg \rho$ which implies  that the warp factor is approximately the AdS one in the UV region, while in IR the conformal symmetry is broken smoothly for $z \simgt 1/\rho$.
The invariant length of the 5th dimension  
which, 
can be approximated as  
\beq
L  = z_0 \left ( \log(1/z_0 \rho) - \gamma/2 \right) + \co (z_0^2).  
\eeq 
The parameter $\rho$ plays a similar role as the IR brane in the hard-wall: 
it makes the invariant length finite and generates a mass gap of order $\rho$.   

We choose the symmetry breaking mass term as 
\beq
M^2(z) = \mu^4 z^2.  
\eeq 
The mass term does not vanish on the UV brane, but $M^2(z_0)$ is suppressed by $z_0^2$ which ensures that the one-loop Higgs mass is not sensitive to the UV scale $1/z_0$. 
The hierarchy problem is avoided when the mass parameter $\mu$ is not much larger than $\tev$. 
The background and the mass term corresponds to the superpotential
\beq
W = {1 \over 2 z} + \sqrt{\mu^4 + \rho^4} \, z , 
\eeq
which can be split as 
\beq
W_0 = {1 \over 2 z} + \rho^2 z  ,
\qquad
U_M = \left ( \sqrt{\mu^4 + \rho^4}  - \rho^2 \right )z . 
\eeq  
Both $W$ and $W_0$ become infinite in IR which shows that the KK spectrum is discrete and has a mass gap.  
$U_M$ fixes the gauge-higgs profile to be
\beq
\eta(z) = \left ( \sqrt{\mu^4 + \rho^4}  - \rho^2 \right )^{-1/2} z_0^{-1/2} 
 \exp\left ( - {1 \over 2} \left ( \sqrt{\mu^4 + \rho^4}  - \rho^2 \right ) (z^2 - z_0^2) \right ),
\eeq 
which is the right half of a gaussian.

In this simple background we can solve the equation of motion explicitly. 
The normalizable solution is 
\beq
K_M(z,p) = e^{- {1 \over 2} \left ( \sqrt{\mu^4 + \rho^4}  - \rho^2 \right )z^2 }
U \left (- {p^2 \over 4 \sqrt{\mu^4 + \rho^4}},0,\sqrt{\mu^4 + \rho^4} z^2  \right ),
 \eeq 
where $U$ is the confluent hypergeometric function of the second kind. 

\begin{figure}[tb]
\begin{center}
\includegraphics[width=0.4\textwidth]{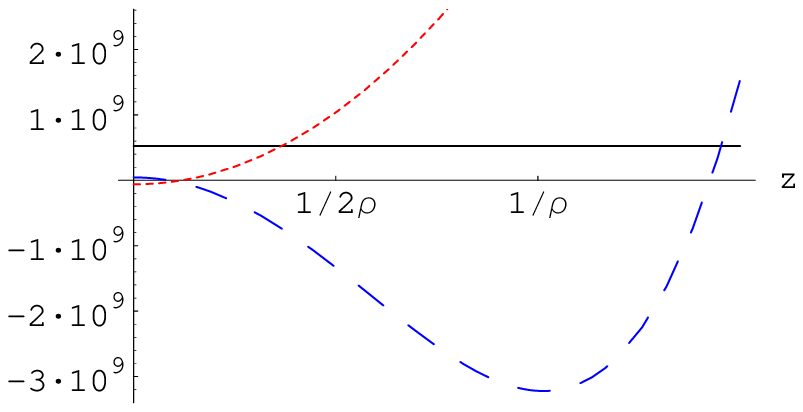}
\includegraphics[width=0.4\textwidth]{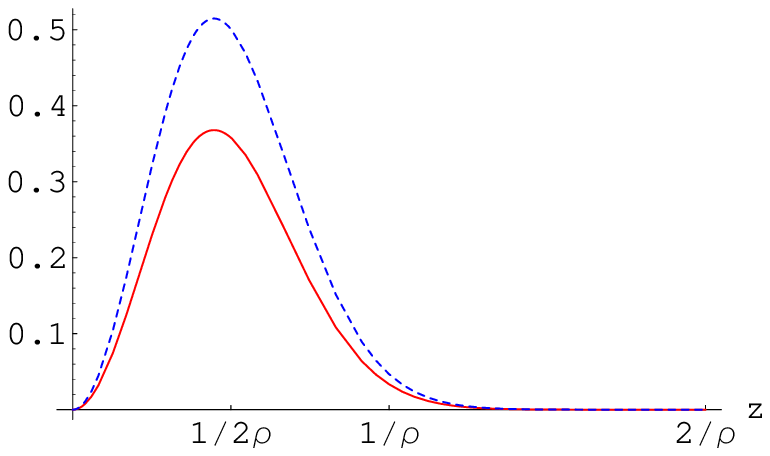}
\caption{Left: 5D profiles of the zero mode and the first two KK modes of the photon in the linear background for $1/z_0 = 10^{19} \gev$, $\rho = 1 \tev$. Right: Embedding of the Higgs boson into the gauge field (solid red) and the bulk scalar (dashed blue) for $\mu = 2.4 \tev$.
} 
\label{f.hp}
\end{center}
\end{figure} 

The photon profile (which does not see the mass term) is proportional to  $K_0(z,p) =  U(- p^2/4\rho^2,0,\rho^2 z^2)$, and its KK spectrum is given by the solutions of $0 = K_0'(z_0,m_n) \to  U(- m_n^2/4\rho^2,1,\rho^2 z_0^2) = 0$. 
To lowest order in $z_0$, the spectrum is  given by the linear Regge trajectory: $m_n \approx 2\rho n^{1/2}$.      
The first KK modes of the W and Z bosons have approximately the same mass, while other vector, scalar and pseudoscalar KK modes are heavier (the splittings depend on the parameter $\mu$).     
Thus, the mass parameter $\rho$ sets the KK scale $\mkk \sim 2\rho$.
Knowing the explicit solution we can find out how the profiles of the KK modes look like, 
and we plotted some examples in fig. \ref{f.hp}. 
Even though the excited KK profiles explode in IR, overlap integrals of the form $\int a(z) [f_n(z)]^i$ are finite thanks to the exponential in the warp factor.

\begin{figure}[tb]
\begin{center}
\includegraphics[width=0.3\textwidth]{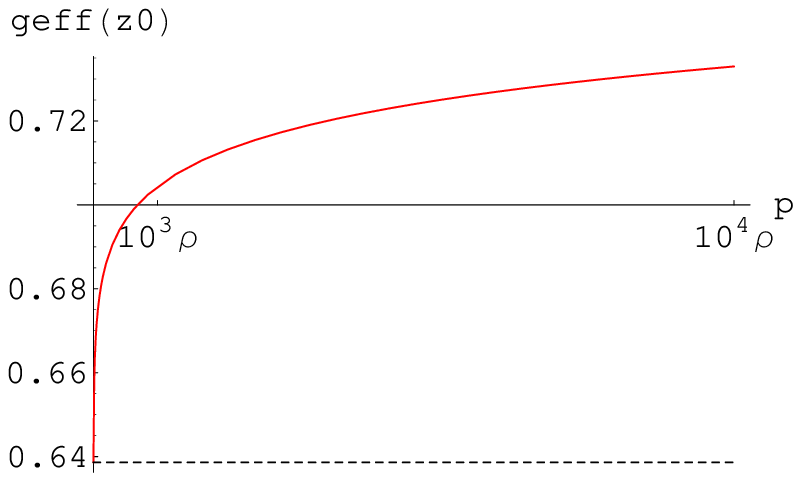}
\includegraphics[width=0.3\textwidth]{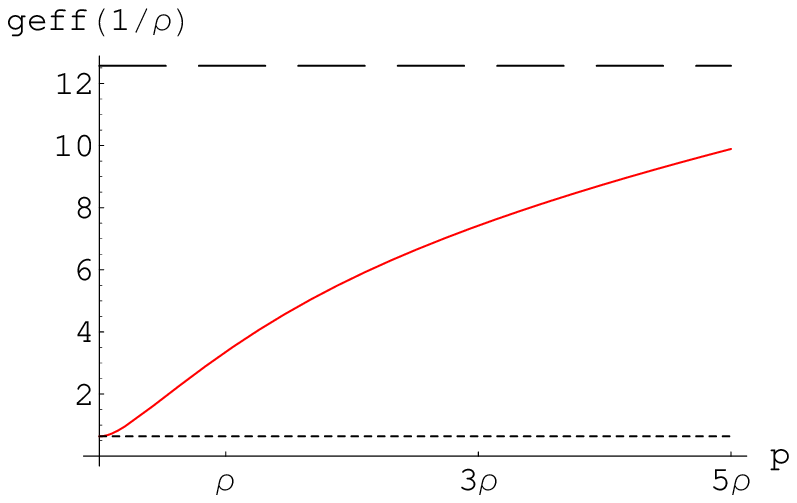}
\includegraphics[width=0.3\textwidth]{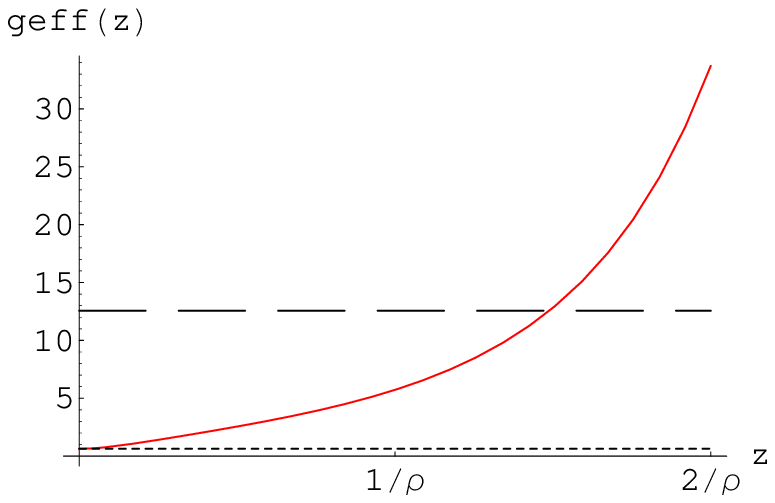}
\caption{The effective  $SU(2)_L$ coupling (solid red) in the linear background ($\rho = 1 \tev$, $1/z_0 = 10^{19} \gev$, $g_* = 3.9$)
as a function of momentum  for $z = z_0$ and $z = 1/\rho$ and as a function of $z$ for $p = 2 \rho$.
The lower dashed line is the SM weak coupling, while the upper dashed line in the second and the third plot marks the strong coupling $g_{eff} = 4 \pi$.} 
\label{f.sc}
\end{center}
\end{figure}

At this point some comments on the perturbativity of our set-up are in order. 
In warped theories, the strong-coupling scale is position dependent. 
One way to quantify it is to introduce the effective coupling 
$g_{eff}^2(z,p) = z_0 g_*^2 p^2 i P(z,z,-p^2)$, where $P$ is the propagator in the 4D momentum/5D position space. 
A physical process involving exchange of KK modes between sources peaked at $z$ is governed by $g_{eff}^2(z,p)$. 
For $p^2 \to 0$ we have $P(z,z,p^2) \to 1/p^2 L$, and the effective coupling approaches the zero mode coupling, independently of $z$.  
While on the UV brane $g_{eff}^2(z_0,p)$ remains perturbative up to very high scales above the mass gap, in the IR $g_{eff}^2(z,p)$ grows as a power of momentum. 
The position dependent strong coupling scale $\Lambda_S(z)$ can be defined as the momentum scale where the effective coupling becomes non-perturbative: $g_{eff}(z,\Lambda_S(z)) \sim 4 \pi$. 
The effective coupling for the electroweak gauge bosons in the linear background is plotted in fig. \ref{f.sc}. 
We can see that on the UV brane the effective coupling grows only logarithmically with momentum while in the IR it grows much faster and quickly hits the non-perturbative values. 
Nevertheless, for  $z = 1/\rho$ the theory includes several KK modes before the strong coupling sets in. 
For $z \gg 1/\rho$, however, the effective coupling becomes non-perturbative below the scale of the first KK mode.  
Thus, sources localized for $z \gg 1/\rho$ (one could, for example, try to localize 3rd generation fermions in the far IR) are unavoidably  strongly coupled in the 5D description.

We move to the electroweak constraints.  
The form factor is given by 
\beq
\Pi_M(p^2)  = - \left ( \sqrt{\mu^4 + \rho^4}  - \rho^2 \right ) z_0 
+ p^2 {z_0 \over 2} 
{U \left (1 - {p^2 \over 4 \sqrt{\mu^4 + \rho^4}},1 ,\sqrt{\mu^4 + \rho^4} z^2  \right ) 
\over 
U \left (- {p^2 \over 4 \sqrt{\mu^4 + \rho^4}},0,\sqrt{\mu^4 + \rho^4} z^2  \right )} .
\eeq 
Since $U(0,b,x) = 1$, $\Pi_M(0)  = - ( \sqrt{\mu^4 + \rho^4}  - \rho^2) z_0$. 
Thus, the global symmetry breaking scale is given by 
\beq
\label{e.fhr} 
f_h^2 =    {2 \over g_*^2}  \left ( \sqrt{\mu^4 + \rho^4}  - \rho^2 \right ) .
\eeq 
The first derivative of the form factor can be approximated as  
\beq
\Pi_M'(0)  = z_0 \left ( - {1 \over 4} \log\left( z_0^4 (\rho^4 + \mu^4) \right) - \gamma/2 \right) + \co (z_0^2) .
\eeq 
Thus, we find the S parameter 
\beq
\label{e.sws} 
S \approx  {\pi v^2 \log \left (1 + {\mu^4 \over \rho^4}\right) \over \sqrt{\mu^4 + \rho^4} - \rho^2} 
= \eps^2 {2 \pi \over g_*^2}  \log \left (1 + {\mu^4 \over \rho^4}\right)  .
\eeq
As opposed to the hard-wall case, the S parameter depends on the combination of  $\rho$ and $\mu$, rather than being directly related to the KK scale.  
Thus, playing with $\mu/\rho$ we are able to relax the 3 TeV bound on the KK scale.   
Comparing \eref{sws} with the hard-wall formula \erefn{hws} we can see that the numerical factor $6$ is replaced on the soft-wall by  $ 2 \log(1 + \mu^4/\rho^4)$. 
Thus, for a fixed $\eps$ and $g_*$, the S parameter can be reduced if $\mu$ is not larger than $\sim 2\rho$.  
 
For the W and Y parameters we find 
\beq
W \approx Y \approx {\pi^2 \over 48} {m_W^2 \over \rho^2 \log (1/z_0 \rho)} . 
\eeq 
As in the hard-wall case, as long as the UV/IR hierarchy is very large, $W$ and $Y$ are suppressed by the log of the hierarchy.
The resulting constraint on the KK scale turns out to be even weaker than in the hard-wall AdS. 

\begin{figure}[tb]
\bc
\includegraphics[width=0.4\textwidth]{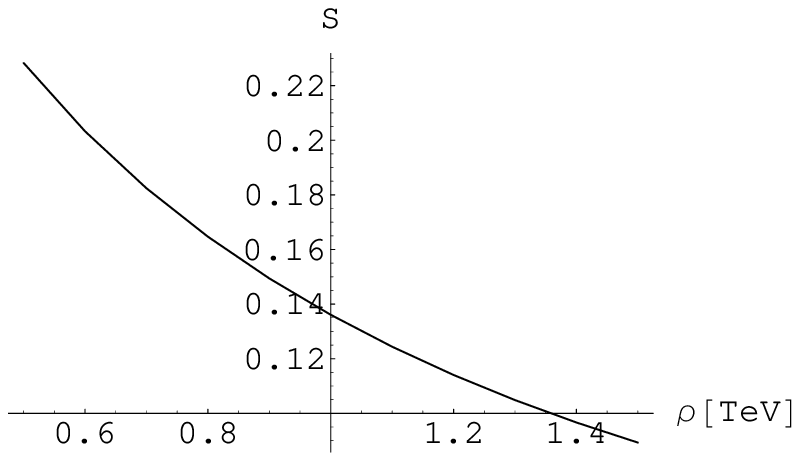}
\includegraphics[width=0.4\textwidth]{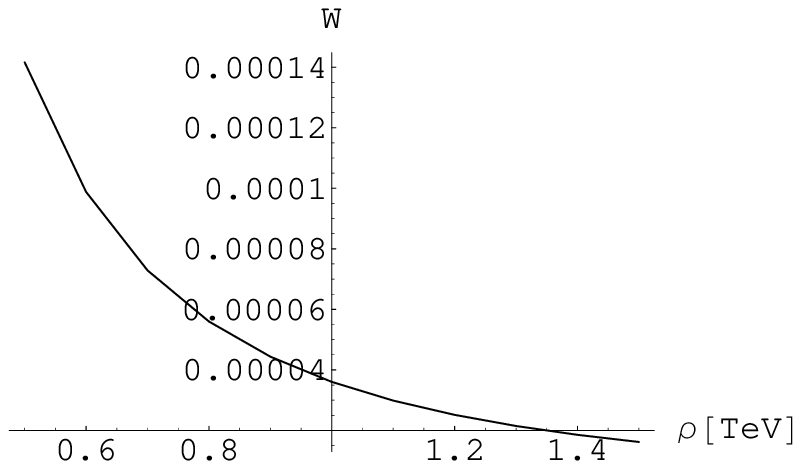}
\caption{S and W parameter in the linear soft wall for $\eps = 0.3$.} 
\label{f.sr}
\end{center}
\end{figure} 
\begin{figure}[tb]
\bc
\includegraphics[width=0.4\textwidth]{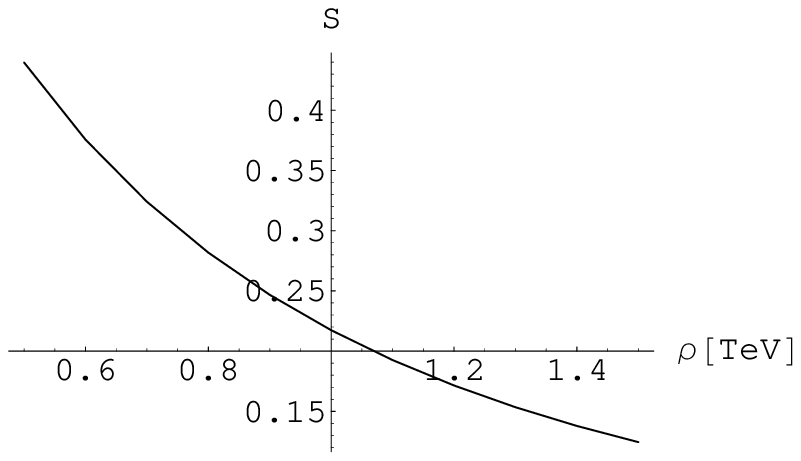}
\includegraphics[width=0.4\textwidth]{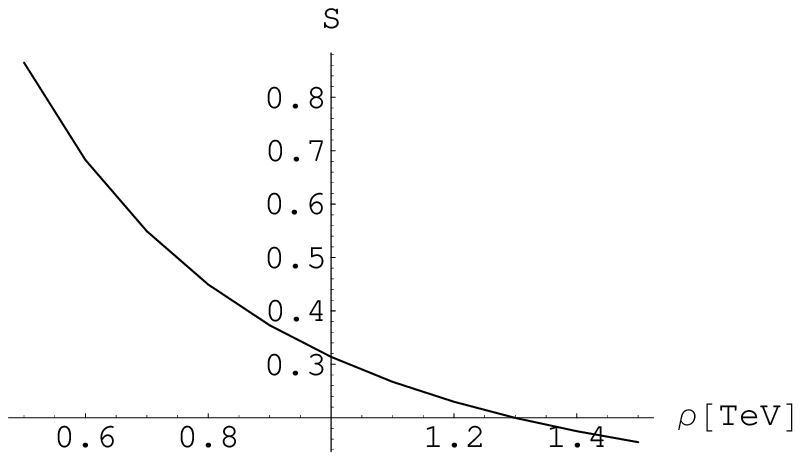}
\caption{S parameter in the linear soft wall for $\eps = 0.5,1$.} 
\label{f.sr2}
\end{center}
\end{figure}

We conclude this analysis with some numerical studies of the electroweak constraints. 
We employ the following procedure 
\ben
\item We fix the UV scale to be of the order of the Planck scale, $1/z_0 = 10^{19}\gev$. 
We also pick up 3 discrete values $v/f_h \equiv \eps = .3,.5,1$. 
We scan over $\rho \in (0.5,1.5) \tev$. 
\item
We assume no UV brane kinetic terms, thus the SM weak coupling is given by $g_L^2 \approx g_*^2 z_0/L$. 
The bulk coupling $g_*$ can be obtained by inverting \eref{fhr}. 
This way, $g_L^2$ becomes a function of $z_0,\rho,\eps,\mu$.
When the first three parameters are fixed,  $\mu$ is determined by matching to the measured weak coupling evaluated at the TeV scale: 
$g_L^2(\tev) \approx 0.41$. 
For our input parameters, we find $\mu \sim 1 \tev$, $g_* \sim 4$.     
\item We plot the S parameter as a function of $\rho$ and find the bounds on the KK scale. 
The results are presented in figs. \ref{f.sr} and \ref{f.sr2}. 
For $\eps = 0.3$, imposing the constraint $S < 0.2$ implies the rather mild bound $\mkk \simgt 1.2 \tev$.
As suggested by \eref{sws}, the constraints become more stringent when we decrease $f_h$. 
For $\eps = 0.5$  we find $\mkk \simgt 2.2 \tev$, while for $\eps = 1$ (the technicolor limit) the bound becomes $\mkk \simgt 2.6 \tev$.  
The W parameter is at most of order $10^{-4}$, safely below the bound $W \simlt 10^{-3}$. 
\een

\subsection{Continuum soft wall}

%
\begin{figure}[tb]
\begin{center}
\includegraphics[width=0.4\textwidth]{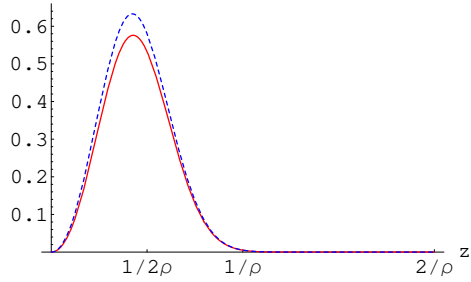}
\caption{Embedding of the Higgs boson into the gauge field (solid red) and the bulk scalar (dashed blue)   in  the continuum background for $\eps = 0.3$, $1/z_0 = 10^{19} \gev$, $\rho = 1 \tev$. 
} 
\label{f.hp2}
\end{center}
\end{figure} 

Our second example has a continuous KK spectrum with a mass gap, which is completely different from anything encountered in the hard wall models.  
The metric is \cite{CMT}
\beq
a(z) = {z_0 \over z} e^{-\rho(z-z_0)},
\qquad 
W_0 = {1 \over 2 z} + {\rho \over 2}.
\eeq
The invariant length of the 5th dimension is 
$L  = z_0 \left ( \log(1/z_0 \rho) - \gamma \right) + \co (z_0^2)$.  
As before, the warp factor is approximately  AdS in UV with conformal invariance broken  at $z \simgt 1/\rho$.
This time, however, the decay of the warp factor in the IR is not fast enough to ensure a discrete spectrum. 
Thus, there will be a continuum of KK modes starting $\rho/2$ (in addition to discrete resonances who feel the $M^2$ term in their Schr\"odinger potential). 


In the continuum case one needs more effort to cook up a tractable example with a sensible symmetry breaking mass term. 
We want the mass to decay in UV, and at the same time we want the potential $V_M$ to be simple enough so that we can find the gauge-higgs profile $\eta(z)$.
The simplest example would be to take $U_M = \mu^2 z$, but then $M^2$ contains a linear term in $z$, which leads to linear sensitivity of the Higgs mass to the UV scale $1/z_0$. 
Therefore we pick up a somewhat more complicated example:  
\beq
U_M = \mu^2 z + \mu^2 \rho z^2 
\qquad \to \qquad M^2(z) = \mu^2 \rho^2 z^2 + \mu^4 z^2 (1 + \rho z)^2 .  
\eeq 
The second term in $U_M$ has been engineered such that  $M^2 \sim z^2$ in UV. 
The gauge-higgs profile is now 
\beq
\eta(z) = {1 \over \sqrt{\mu^2 z_0 (1 +\rho z_0 )}}  
 \exp\left ( - {1 \over 2}  \mu^2(z^2 - z_0^2) - {1 \over 3}  \mu^2 \rho(z^3 - z_0^3) \right )
\eeq 
and the global symmetry breaking scale is fixed by $\mu$, 
\beq
\label{e.fhc} 
f_h^2 =    {2 \over g_*^2} \mu^2 \left (1 + \rho z_0 \right ) .  
\eeq

\begin{figure}[tb]
\bc
\includegraphics[width=0.4\textwidth]{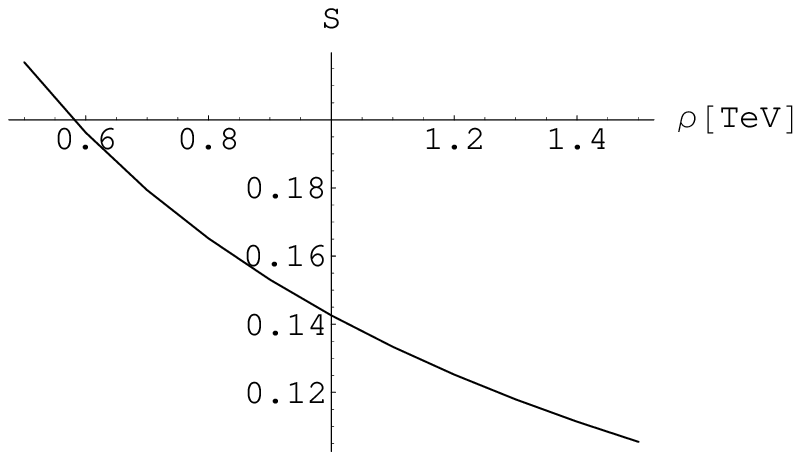}
\includegraphics[width=0.4\textwidth]{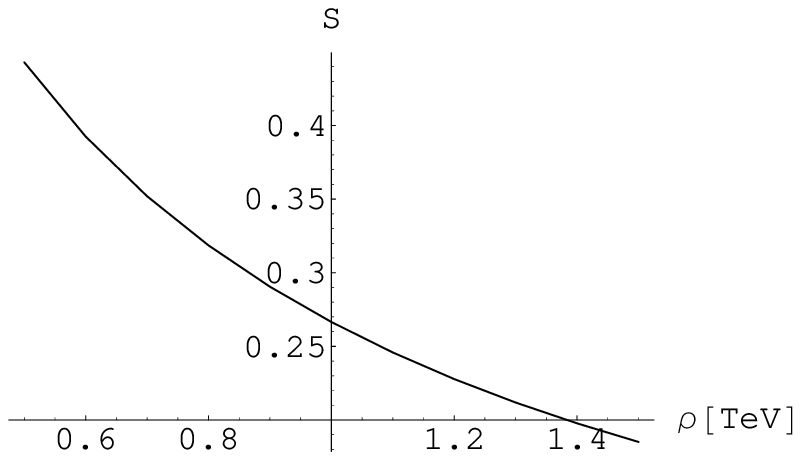}
\caption{S parameter in the continuum soft wall for $\eps = 0.3,0.5$.} 
\label{f.hws}
\end{center}
\end{figure}

This time our task is a bit harder, as we are not able to  solve the equations of motion in this background.
At this point we should appreciate the formulas that express the oblique parameters as integrals of the warp factor.  
We follow the same procedure as in the linear case, assuming no UV kinetic terms and fixing $\mu$ to match the weak gauge coupling. 
Lacking analytical results, we obtain the S parameter by evaluating \eref{asd} numerically.
The results in fig. \ref{f.hws} show that the possibility of  a continuum of KK modes is surprisingly weakly constrained by the electroweak precision data. 
Imposing $S < .2$ constrains $\rho < 0.6 \tev (1.4 \tev)$ for $\eps = 0.3 (0.5)$.
Given that continuum starts at $\rho/2$, in both cases new physics below $1 \tev$ is perfectly compatible with the experimentally observed smallness of the S parameter.      

\section{Summary and Outlook}
\label{sec:d} \setcounter{equation}{0} \setcounter{footnote}{0}

We extended the formulation of 5D gauge-Higgs unification to the soft wall framework where the IR brane is replaced by a smoothly decaying warp factor.
5D gauge symmetry is broken by UV boundary conditions and by a vev of a bulk scalar field. 
The Higgs boson lives partly in the 5th component of the gauge field, and partly along the Goldstone direction of the bulk scalar.   
The soft-wall version can maintain the attractive feature of the standard gauge-Higgs scenario that the loop induced Higgs potential does not suffer from the large hierarchy problem. 
More precisely, the Higgs potential is insensitive to the scale of the UV  brane is if the bulk scalar condensate is localized in IR and vanishes as $z^2$ or faster in UV.   

We argue that our construction is more than a formal exercise. 
Soft wall is a box of new possibilities in KK phenomenology, allowing for new kinds of spectra and couplings of the KK resonances.  
One can even construct phenomenologically viable examples where the KK spectrum is continuous above a mass gap, with potentially striking hidden-valley phenomenology. 
Most interestingly, bounds from electroweak precision test that create some tension in the standard hard-wall scenario can be relaxed. 
We presented one explicit example where the bound from the S parameter on the lightest KK gauge boson mass is 2 TeV, rather than 3 TeV for the hard-wall. 
Somewhat surprisingly, the electroweak constraints on the hidden-valley-type spectra turn out  be even weaker than the constraints on discrete resonances, allowing for a continuum starting below 1 TeV.   
Softer is often safer. 

Focused on the low energy phenomenology of the gauge sector, we left a couple of loose ends on the model-building side.  
Firstly, bulk fermions were not included.
Because of that, we did not touch the flavor issues that usually are realized by different wave function localization of the SM fermions.   
Moreover, we could not compute the fermion contribution to the radiative Higgs potential.  
Since gauge fields yield a  positive contribution to the Higgs mass squared, we simply assumed that the fermion contribution is negative and of appropriate magnitude to arrive at an electroweak breaking vacuum with $v/f_h < 1$.  
It would be interesting to have an explicit realization of the fermion sector to see if the soft-wall scenario allows us to reduce the fine-tuning of electroweak breaking.    
Secondly,  we did not obtain our soft wall backgrounds as solutions of the equations of motion. 
That would allow us to address the issues of  back-reaction of the scalar condensate, radion stabilization and so on. 
We restrained ourselves from solving all these problems here, so as to leave some for future publications.

\section*{Acknowledgements}

We would like to thank Csaba Csaki and Tony Gherghetta for important discussions. 
A.F. is ever so grateful to Francesco Riva for the Euro'08 tickets. 
A.F. is partially supported by the European Community Contract MRTN-CT-2004-503369 for the years 2004--2008. 
M.P.V. is supported in part by MEC project FPA2006-05294 and Junta de Andaluc\'{\i}a projects FQM 101, FQM 00437 and FQM 03048.

\appendix

\section*{Appendix}

\section{Holographic Derivation of the Low-Energy Effective Action}
\setcounter{equation}{0}
\label{a.hd}

The cleanest way to compute the higher order corrections to the SM lagrangian in the universal version of the 5D setup  is by using the holographic approach \cite{BPR}.  
It consists in choosing the UV boundary value of the bulk gauge fields (rather than the zero modes) as the dynamical variable in the low energy effective theory, while the bulk degrees of freedom are integrated out.
In the following we derive the effective action using the language of the propagators in the mixed 4D momentum/5D position space (p/z propagators, in short).    
To this end  we first rewrite the quadratic action in the Fourier-transformed basis:   
\beq
S = \int {d^4 p \over (2 \pi)^4} \int_{z_0}^\infty dz  {1 \over 2}  A_\mu(p,z) D_{\mu \nu}(p,z)  A_\nu(p,z) , 
\eeq  
where $D_{\mu\nu}$ is the kinetic operator,  
\beq
D_{\mu\nu} = a(z) \left [ - \eta_{\mu \nu} p^2 + p_{\mu} p_\nu (1 - 1/\xi) \right] I  + \eta_{\mu \nu} D_z (a(z) D_z)
\eeq 
The boundary terms from integration by parts vanish due to the boundary conditions and we have added an $R_\xi$ gauge fixing term. 
The p/z propagator is defined as the inverse of the kinetic operator: 
\beq
D_{\mu\rho}(p,z) P_{\rho\nu}(p,z,w) = i \delta(z-w) \eta_{\mu \nu} I .
\eeq 
$P$ is a matrix in the $SO(5)\times U(1)_X$ group space, and $I$ is the identity matrix in that space.  
The propagator satisfies the same boundary conditions as the gauge field.
On the UV brane, at $z = z_0$,
\beq 
\pa_z P_{L_a A} = P_{R_i A} = P_{C_{\hat a} A} = 0 ,
\qquad 
\pa_z (s_x P_{R_3 A} + c_x P_{X A}) =c_x P_{R_3 A} - s_x P_{X A} = 0 ,
\eeq
while the conditions in IR are that propagators should be expressed by normalizable solutions to the equations of motion.   

In our formalism, the effective action is obtained by writing the 5D gauge field as   
\beq
A_\mu(p,z) = P_{\mu \nu} (p,z,z_0) \bar A_\mu(p).   
\eeq  
and plugging this back into the 5D action.
This integrates out, at the tree level, the bulk degrees of freedom, leaving the boundary source $\bar A_\mu(p)$ as the dynamical variable of the low energy effective theory.  
The quadratic terms in the effective action are given by the UV boundary propagator:  
\beq
\cl_{eff} = \bar A_\mu(p) P_{\mu \nu} (p,z_0,z_0)  \bar A_\nu(p). 
\eeq 
In the following we choose $\xi = 1$, which implies that $P_{\mu\nu}(p,z,w) = \eta_{\mu\nu}P (p^2,z,w)$, and we will  denote the boundary value $P (p^2,z_0,z_0)$ as  $\bar P$. 
Note that $P (p^2,z,w)$ satisfies the equations of motion  $(a^{-1} D_z (a D_z) + p^2) P(p^2,z,w)  = i \delta (z - w)$ that, for $z \neq w$ reduces to the usual equation of motion in the gauge-higgs background.   
The propagator is computed along the same lines as solving the equations of motion, which we described in \eref{wr} and below. 
After some algebra we find that  can be expressed in terms of the form factor $\Pi_M(p^2)$ which, in turn  can be written in terms of the normalizable solutions to the equations of motion:
\beq
\label{e.ffa}
\Pi_M(p^2) = \pa_z \log \left (K_M(z,p) \right)|_{z = z_0} .
\eeq  
Explicitly, the non-vanishing boundary propagators are given by 
\bea
i \bar P_{L_i L_i} &=& 
{1 \over \Pi_0(p^2) + {\sin^2(\ti v/f_h) \over 2} \left (\Pi_M(p^2) -\Pi_0(p^2)\right ) },
\nn
i \bar P_{L_3 L_3} &=& 
{1 + s_x^2 {\sin^2(\ti v/f_h) \over 2} {\Pi_M(p^2) -\Pi_0(p^2)  \over \Pi_0(p^2)} \over 
\Pi_0(p^2) +  (1 + s_x^2){\sin^2(\ti v/f_h) \over 2} \left (\Pi_M(p^2) -\Pi_0(p^2) \right ) } ,
\nn
i \bar P_{L_3 B} &=& 
 {s_x {\sin^2(\ti v/f_h) \over 2} {\Pi_M(p^2) -\Pi_0(p^2)  \over \Pi_0(p^2)}   \over 
\Pi_0(p^2) +  (1 + s_x^2){\sin^2(\ti v/f_h) \over 2} \left (\Pi_M(p^2) -\Pi_0(p^2)\right )  } ,
\nn
i \bar P_{B B} &=& 
{1 +   {\sin^2(\ti v/f_h) \over 2} {\Pi_M(p^2) -\Pi_0(p^2)  \over \Pi_0(p^2)}   \over 
\Pi_0(p^2) +  (1 + s_x^2){\sin^2(\ti v/f_h) \over 2} \left (\Pi_M(p^2) -\Pi_0(p^2)\right )   } ,
\eea 
where $\bar P_{L_3 B} = \bar P_{L_3 R_3}/s_x = \bar P_{L_3 X}/c_x$ and 
$\bar P_{B B} = \bar P_{R_3 R_3}/s_x^2 = \bar P_{X X}/c_x^2 = \bar P_{R_3 X}/c_x s_x$.  

Next, we generate the SM electroweak bosons $L_\mu^a$, $B_\mu$ from the boundary sources $\bar A(p)$ in the following way: 
\bea
L_\mu^{i} &=& Z_L^{1/2} \bar P_{L_i L_i} \bar L_\mu^i ,
\nn
L_\mu^{3} &=& Z_L^{1/2} \left [ \bar P_{L_3 L_3} \bar L_\mu^3  +   \bar P_{L_3 B}(s_x \bar R_\mu^3 + c_x \bar X_\mu) \right ] ,
\nn
B_\mu  &=&  Z_B^{1/2} \left [ \bar P_{L_3 B} \bar L_\mu^3  +    \bar P_{B B}(s_x \bar R_\mu^3 + c_x \bar X_\mu) \right ] .
\eea 
$Z_{L,B}$ are momentum independent normalization factors that will be chosen such as to make the kinetic terms canonically normalized.  
This particular choice of basis ensures that the interactions of the UV boundary fermions with $L_\mu^a$, $B_\mu$  have the SM form:  $g_L \ov \psi L^a T_L^a \psi + g_Y \ov \psi B Y \psi$, with $g_L = Z_L^{-1/2} g_5$ and $g_Y = Z_B^{-1/2} s_x g_5$. 
Hence, in this basis there is no tree-level vertex correction whatsoever.
Four-fermion terms are not generated either because,  by locality, UV boundary fermions couple only to the boundary gauge bosons. 
The corrections to the SM lagrangian show up only in the electroweak gauge boson sector. 
Quite concisely, the quadratic terms can be given in term of the two form factors defined before:  
\begin{align}
\label{e.gela} 
\cl_{eff} = &  - {1 \over 2} \Pi_0(p^2) \left [ 
Z_{L}^{-1} L_\mu^a L_\mu^a   + Z_{B}^{-1} B_\mu B_\mu \right ] 
\nn
& \mbox{} + Z_{L}^{-1} {\sin^2(\ti v/f_h) \over 4} \left (\Pi_0(p^2) -\Pi_M(p^2) \right )
\left [ 
L_\mu^i L_\mu^i + (L_\mu^3 - {g_Y \over g_L} B_\mu )^2 \right ] .
\end{align}


\end{document}